\documentclass[12pt,letterpaper]{JHEP3}

\title{Gauged Gravity via Spectral Asymptotics
of non-Laplace type Operators}

\author{Ivan G. Avramidi\\
Department of Mathematics,
New Mexico Institute of Mining and Technology,
Socorro, NM 87801, USA\\
E-mail: \email{iavramid@nmt.edu}}

\abstract{We construct invariant differential operators acting on 
sections of vector bundles of densities over a smooth manifold
without using a Riemannian metric. The spectral invariants of such
operators are invariant under both the diffeomorphisms and the gauge 
transformations and can be used to induce a new theory of gravitation.
It can be viewed as a matrix generalization of Einstein general relativity
that reproduces the standard Einstein theory in the weak deformation limit.
Relations with various mathematical constructions such as Finsler geometry
and Hodge-de Rham theory are discussed.}


\preprint{New Mexico Tech, May 28, 2004}

\def\nmtstamp#1
{\null\par\vskip-3truecm\par\hfill
\begin{minipage}{90mm}
{\hrulefill}\par\vskip-4truemm\par
{\hrulefill}\par\vskip5mm\par
{{\large\sc New Mexico Tech} \large\textrm{(#1)}}
\vskip4mm\par{\hrulefill}\par\vskip-4truemm\par
{\hrulefill}\par
\end{minipage}
} 
\def\sideremark#1
{\ifvmode\leavevmode\fi\vadjust{\vbox to0pt{\vss
\hbox to 0pt{\hskip\hsize\hskip1em
\vbox{\hsize2cm\tiny\raggedright\pretolerance10000
\noindent #1\hfill}\hss}\vbox to8pt{\vfil}\vss}}}

\def\Alt{\mathrm{\,Alt\,}}
\def\Aut{\mathrm{\,Aut\,}}
\def\CC{{\mathbb C}}

\def\det{\mathrm{\,det\,}}
\def\Det{\mathrm{\,Det\,}}

\def\End{\mathrm{\,End\,}}

\def\II{{\mathbb I}}

\def\Id{\mathrm{\,Id\,}}

\def\log{\mathrm{\,log\,}}
\def\Mat{\mathrm{\,Mat\,}}

\def\RR{{\mathbb R}}

\def\tr{\mathrm{\,tr\,}}
\def\Tr{\mathrm{\,Tr\,}}

\def\be{\begin{equation}}
\def\ee{\end{equation}}
\def\bea{\begin{eqnarray}}
\def\eea{\end{eqnarray}}
\def\bes{\begin{displaymath}}
\def\ees{\end{displaymath}}
\def\bmp{\begin{minipage}}
\def\emp{\end{minipage}}
\newtheorem{theorem}{Theorem}

\begin{document}

\section{Introduction}

Einstein General  Relativity is accepted as a correct theory of  gravity
at huge range of scales, from cosmological to the subatomic ones, that  
successfully  describes almost all classical gravitational phenomena (with
few exceptions like singularities, dark energy, etc). In spite of
this,  there is no  consistent theory of quantum gravitational
phenomena, that is of physics at very small length scales. It is
expected that the general relativistic  description of gravity, and, as
the result, of the space-time, is inadequate at short distances
(and maybe at very large distances due to the cosmological constant 
problem). There
are many different proposals how to extend General Relativity.  Aside
from the well established approaches such as string theory
and loop gravity 
some work has
been done in noncommutative theory of gravity associated with
noncommutative spaces
\cite{chamseddine93a,chamseddine93b,chamseddine03,madore94}
as well as ``matrix gravity'' 
\cite{okubo85,maluf88,mann84,cutler87,wald87,avramidi03a, avramidi03b,
anco03,chamseddine04}.

One of the first attempts of constructing matrix gravity was proposed by
Okubo and Maluf \cite{okubo85}, whose primary goal was to unify
Einstein's general relativity and Yang-Mills gauge theory. The main
ingredient in their approach was the matrix valued  affine connection
and the main purpose of their work was the unification of Einstein
general relativity and Yang-Mills theory. They introduced the
matrix-valued metric (or the frame vectors) and the connection
coefficients and developed the formalism of some kind of 
``matrix-valued'' differential geometry, that is, covariant derivatives,
curvatures etc. Because of the non-commutativity many choices have to be
made in any ``matrix'' differential geometry, such as how to raise
indices, how to define covariant derivatives, what constraints to use to 
fix the torsion and some others. Different choices lead to different 
physical theories and there is no physical principle which could help in
distinguishing a unique possibility. The problems encountered by Okubo
and Maluf were: the presence of the spin-3 components of the connection,
non-uniqueness of connection, non-covariance of the torsion,
non-covariance of higher covariant derivatives, consistency of covariant
derivatives of higher rank tensors and some others. The further work in
this direction \cite{maluf88} was focused on an particular ansatz for
the affine connection which was introduced originally by Einstein and
Kaufmann \cite{einstein55} in the Abelian theory. This ansatz separates
the usual Christoffel coefficients and the Yang-Mills gauge fields and
leads to a standard Einstein general relativity  coupled to the
Yang-Mills theory with a particular gauge group.  There are no new
physical degrees of freedom in this formulation.

The formalism of matrix geometry is related to the algebra-valued
tangent space formulation of Mann \cite{mann84} and Wald \cite{wald87}.
In particular, Wald introduced the notion of an algebra-valued tangent
space (more generally, algebra-valued tensor fields) and  generalized
the whole formalism of differential geometry  (that is covariant
derivatives, metrics, forms  and curvature) to the algebra-valued case.
This geometric interpretation was motivated by the preceding work of
Cutler and Wald \cite{cutler87}, where  they found a {\it new type of
gauge invariance} for a collection of {\it massless} spin-2 fields, more
precisely, a consistent theory describing the  interaction of a
collection of massless spin-2 fields. The infinitesimal form of the new
gauge transformations of the algebra-valued metric  looks precisely like
the infinitesimal diffeomorphisms but with {\it algebra-valued
infinitesimal vector fields}. The consistency condition simply says that
the commutator of two  infinitesimal gauge transformations is an
infinitesimal gauge transformation. The main conclusion of  Cutler and
Wald was that the  consistency  conditions can only be satisfied for
{\it associative commutative algebras}. Therefore, all constructions
``diagonalize'' (everything commutes) and the interacting theory of a
collection of massless spin-2 fields is simply a sum of the usual
Einsten-Hilbert actions for each field (no cross-interactions). This
restricts significantly the possible physical applications of this
approach. Further, Wald noticed that the algebra-valued vector fields do
not generate diffeomorphisms of a real manifold. Instead, they are
diffeomorphisms on so-called `algebra manifold', in which the {\it
coordinates themselves are algebra-valued}. The supersymmetric extension
of this approach was continued in 
\cite{anco03}.

Similar results concerning consistent cross-interactions of a collection
of {\it massless} spin-2 fields were obtained in \cite{boulanger00},
where the multi-graviton theories were analyzed (from a different point
of view). The authors of this work considered a collection of {\it
massless} spin-2 tensor fields such that:  i) the Lagrangian contains no
more than two derivatives of the fields, ii) the interactions can be
continuously switched on, and iii) in the limit of no interaction, the
Lagrangian reduces to the sum of Pauli-Fierz Lagrangians for each field.
The authors of this paper consider the  free limit condition described
above to be the crucial one. They  mention that Cutler and Wald did not
analyze the extra conditions that must be imposed on the gauge
symmetries coming from the free limit condition. They indicate that none
of the examples of Cutler and Wald models have the correct free limit,
since some of the gravitons come with the wrong sign and, therefore, the
energy of the theory is unbounded from below.  The free limit
requirement gives an extra constraint, which implies that the algebra is
not only commutative and associate but also  {\it symmetric}. Therefore,
the algebra has a trivial structure: it is simply the direct sum of
one-dimensional ideals. This eliminates all cross-interactions. The only
consistent deformation of the sum of Pauli-Fierz Lagrangians is the sum
of Einstein-Hilbert Lagrangians (with a cosmological term) for each
field.

In our previous papers  \cite{avramidi03a,avramidi03b} we proposed to
describe gravity by a multiplet of tensor fields   with the
corresponding gauge symmetry incorporated in the model. We would like to
stress from the very beginning that our approach   is different from the
schemes studied by previous authors. Our main ingredient is the  {\it
matrix-valued two-tensor field}, so that the components of these tensor
fields  {\it do not commute} with each other, in general. Our algebra is
associative but {\it noncommutative}. The other difference (related to
the first one) is the form of the gauge transformations. We start from
the very beginning with a real manifold with real-valued coordinates and
the usual diffeomorphism group, so that there is no problem of defining
the finite gauge transformations.  Our gauge group is  simply the
product of diffeomorphisms and an internal group, say $U(N)$. That is
why we do not have the inconsistences studied by \cite{cutler87,wald87}
and
\cite{boulanger00} and we can allow our algebra to be non-commutative. 
In our approach there is a collection of tensor spin-2 fields, but  {\it
only one of them is massless}, all other fields are massive.  How
exactly this happens depends on the details of the symmetry breaking
mechanism etc, but since we have only one (the usual real-valued) 
diffeomorphism group, only one field is massless. The parameter of our
gauge transformations is a {\it real valued vector field}, not the
algebra-valued vector fields of \cite{cutler87,wald87,boulanger00}
needed to describe multiple {\it massless} spin-2 fields, which
transform {\it independently} of each other.
A similar approach to matrix gravity was proposed recently by 
Chamseddine \cite{chamseddine04}, who considered an $SL(2N,\CC)\otimes
SL(2N,\CC)$ gauge theory with a spontaneous breakdown of symmetry 
giving one massless spin-2 field and $(2N^2-1)$ massive spin-2 fields.

Our idea was simply to 
repeat the Einstein's analysis of the causal structure of space-time.
We showed that the basic notions of general relativity are
based on the geometrical interpretation of the wave equation that
describes propagation of fields {\it without internal structure}, (in
particular, light), that could  transmit information in the spacetime.
At the microscopic distances this role of the electromagnetic field
could be played by other fields {\it with some internal structure}. That
is why instead of a scalar wave equation we studied a {\it hyperbolic
system} of linear second-order partial differential equations. This
cardinally changes the standard geometric interpretation of General
Relativity.  Exactly in the same way as a scalar equation defines
Riemannian geometry, a system of wave equations  generates a more
general picture, that we call {\it Matrix Geometry}.

Morevover, we developed some
kind of ``matrix geometry". We defined the matrix-valued metric, 
affine connection, torsion and curvature. We have also found a 
particular form of a compatibility condition that enabled us to 
{\it explicitly} express the connection in terms of the derivatives 
of the metric. By using the matrix curvature
we constructed an action functional that:
i) contains no more than two derivatives of the fields in such a way
that the equations of motion contain no more than two derivatives,
ii) is invariant under both diffeomorphisms and the gauge transformations,
iii) reduces to the sum of the Einstein-Hilbert action and 
the Yang-Mills action in the weak deformation limit.
Notice that we do not obtain multiple Einstein-Hilbert actions, 
but just one.
As we mentioned above there are several choices to make when
generalizing the real-valued geometry to the matrix-valued one. In
particular, the definition of the ``measure'', raising and lowering
indices, definition of higher-order  covariant derivatives, torsion
constraints, and some others. As a result the action constructed in
these papers is {\it not unique}. 

In the present paper we propose an additional {\it physical principle}
({\it Low Energy Limit}) to construct an action functional that has all
of the above properties but is also {\it unique} (or natural). We simply
require that {\it the classical action of gravity is the semiclassical
limit of the effective action of matter fields interacting with the
gravitational field}. This is nothing but the ideology of ``induced
gravity''. It is well known that  the usual Einstein-Hilbert action
(with a cosmological term) is just the first two terms of the asymptotic
expansion of the effective action, or, simply, the first two
coefficients of the asymptotic expansion of the trace of the heat kernel
for a certain invariant second-order self-adjoint elliptic (in Euclidean
formulation) partial differential operator with a {\it scalar} leading
symbol given by the Riemannian  metric. That is why, in the present
paper we propose to define the action of matrix gravity in a similar way
as a linear combination of the first two coefficients of the trace of
the heat kernel for a more general differential operator, with a {\it
matrix-valued leading symbol} given by the matrix-valued metric.  This
way we avoid the constructions of matrix geometry, such as curvatures
etc, which are non-unique. Although the language of differential
geometry is certainly helpful, we simply do not need it here. In this
approach {\it we get the invariants not from the curvatures but from the 
spectral invariants of a differential operator with a matrix-valued
leading symbol}. This is the main idea  and the main difference of the
present paper from our previous papers
\cite{avramidi03a,avramidi03b}. 

%
%

The outline of the present paper is as follows. In Sect. 2.1 we
introduce the necessary differential-geometric framework. We define the
vector space of complex $N$-vectors and $N\times N$ matrices. In Sect.
2.2 we introduce the fields as {\it densities}. This is done to avoid
the definition of the measure, which we do not have since we do not have
the metric. In Sect. 2.3 the matrix-valued metric and the matrix-valued
Hamiltonian are introduced and certain non-degeneracy conditions are
imposed. We consider two cases: elliptic and hyperbolic metric,
depending on the eigenvalues of the  Hamiltonian. In Sect. 2.4 it is
shown that the above construction is closely related to a collection of 
Finsler geometries and the Finsler metrics are defined. In Sect. 2.5 we
consider matrix-valued $p$-forms and define the matrix-valued star
operator $*$ (similar to the Hodge star operator). In order to define
this operator in an invariant way we also introduce an additional
ingredient: a matrix-valued measure. However, this measure is not
related to the matrix-valued  metric in the usual way (since it is not
unique in the matrix case).  Rather it is considered as an additional 
degree of freedom.

In Sect. 3 we define the invariant differential operators  of the first
and the second order acting on matrix-valued $p$-forms. We define the 
exterior derivative, the coderivative as well as their covariant
versions in Sect. 3.1 and the gauge curvature in Sect 3.2. By using the
covariant derivative and the coderivative we define invariant
second-order differential operators (Laplacians) in Sect. 3.3 and
invariant first-order differential operators (Dirac operators) in Sect
3.4. In Sect. 3.5 we present the main operator studied in this paper in
local coordinates. This is used later in the paper.

In Sect. 4 we list several particular  examples of decomposition of the
general matrix valued metric and introduce the deformation parameter. 
We consider both commutative and noncommutative cases and compare them
with the approaches of previous authors.

In Sect. 5 we briefly summarize the heat kernel approach to compute the
spectral asymptotics, the zeta-function regularization, and the
asymptotic expansion of the effective action. The heat kernel
coefficients introduced here play the crucial role in future
development. In Sect. 5.1 we compute the first two coefficients of the
asymptotic expansion of the trace of the heat kernel for the non-Laplace
type differential operator of general type (that is, a second-order
operator with a non-scalar leading symbol). The Theorem 1 is one of the
main results of  the present paper. In Sect. 5.2 we use the results of
the Theorem 1 to calculate the coefficients $A_0$ and $A_1$ in the
commutative (scalar) case and confirm the well known results. This
serves as a `consistency check' of the calculations of Sect. 5.1 as well
as shows how to make use of the general formulas of Theorem 1.

In Sect. 6 the results for the heat kernel asymptotics are used to construct
the action of ``matrix gravity" via the induced gravity approach.
The form of the action in the commutative limit is provided. It contains
a metric and a scalar field (coming from the additional degree of freedom
associated with the measure).

In Sect. 7 the results are briefly discussed and a list of open 
interesting problems (singularities, high-energy behavior, dark energy,
confinement etc) is
presented.  

\section{Vector Bundles}

\subsection{Vector Spaces}

To be precise, let $V$ be a $N$-dimensional complex  vector space with a
Hermitian inner product $\left<\;,\,\right>$.  As usual, the dual vector
space $V^*$ is naturally identified with $V$, $\varphi \mapsto
\bar\varphi$, by using the inner product: 
\be
\left<\psi,\varphi\right>=\bar\psi(\varphi)\,,
\ee
where $\varphi,\psi\in V$ and $\bar\psi \in V^*$.
If the elements $\varphi$ of the vector space $V$ are
represented by $N$-dimensional contravariant vectors
$\varphi=\left(\varphi^A\right)$, $(A=1,\dots,N)$,
and the elements of the dual space $V^*$ by covariant vectors
$\bar\psi=(\bar\psi_B)$,
then the inner product is defined via the Hermitian metric,
$E=(E_{\dot A B})$,
\be
\left<\psi,\varphi\right>=\psi^{\dagger}E\varphi
=\psi^{A}{}^{*}E_{\dot A B}\varphi^B=\bar\psi_B\varphi^B\,,
\ee
where
\be
E^{\dagger}=E\,,\qquad
\bar\psi=\psi^\dagger E\,,
\ee
or
\be
\bar\psi_B=\psi^A{}^*E_{\dot A B}=(E_{\dot B A}\psi^A)^*\,.
\ee
Here and everywhere below we use the Einstein summation convention and
follow the standard notation to mark the indices of the
complex conjugated fields by dots; the symbol $*$ denotes the
complex conjugation and the symbol $\dagger$ denotes the
Hermitian conjugation.

The vector space $\End(V)$ of endomorphisms of the vector 
space $V$ is isomorphic to the vector space $\Mat(N,\CC)$ of complex 
$N\times N$ matrices $X=(X^A{}_B)$. 
The group of automorphisms $\Aut(V)$ of the vector space $V$ is
isomorphic to the general linear group $GL(N,\CC)$ of complex
nondegenerate $N\times N$ matrices.
The adjoint $\bar X$ of 
an endomorphism $X$ is defined by
\be
\bar X=E^{-1} X^\dagger E\,,
\ee
so that for any $\varphi,\psi\in V$
\be
\left<\psi,X\varphi\right>=\left<\bar X\psi,\varphi\right>\,.
\ee

The metric determines the subgroup of unitary endomorphisms
(or isometries) $G$ of the group of the   automorphisms 
$\Aut(V)$
of the vector space $V$ preserving the inner product, that is 
\be
G=\left\{U\in GL(N,\CC)\;|\;\bar U U=\II\right\}\,, 
\ee
where $\II$ is the identity endomorphism. The dimension of the group $G$ is
$\dim G=N^2$. In the case $E=\II$ the group $G$ is nothing but 
$U(N)$.

\subsection{Vector Bundles of Densities}

Now, let $M$ be a smooth compact orientable 
$n$-dimensional manifold without
boundary and 
${\cal V}\left[w\right]$
be a smooth  vector bundle  of
{\it densities} of weight $w$ over the manifold $M$
with the fiber $V$. 
Here and everywhere below we indicate explicitly
the weight in the notation of the vector bundles.
That is, the sections $\varphi$ of the vector bundle 
${\cal V}[w]$
are vector-valued functions $\varphi(x)=(\varphi^A(x))$ 
that transform 
under the diffeomorphisms $x'^\mu=x'^\mu(x)$ according to
\be
\varphi'(x')=J^{-w}(x)\varphi(x)\,,
\ee
where 
\be
J(x)=\det\left[{\partial x'^\mu(x)\over \partial x^\alpha}\right].
\ee
Hereafter we label the local coordinates on the manifold $M$ by Greek
indices which run over $0,1,\dots,n-1$. 
The small Latin indices will be used to label the
space coordinates, i.e. the local coordinates on hypersurfaces transversal
to the time coordinate; they will run over $1,2,\dots,n-1$.

Further, we denote by ${\cal V}^*\left[w\right]$ the 
dual bundle of densities of weight $w$
with the fiber $V^*$
and by $\End({\cal V})[w]$ the bundle of endomorphisms-valued 
densities of weight $w$ of the bundle ${\cal V}\left[w\right]$
with elements being matrix-valued functions
$U=(U^A{}_B(x))$.

We will also consider the bundle of vector-valued
antisymmetric densities of weight $w$ of type  $(0,p)$ ($p$-forms),
$\Lambda_p[w]=\left((\wedge^p T^*M)\otimes {\cal V}\right)[w]$, 
and the bundle of anti-symmetric densities of weight $w$ 
of type $(p,0)$,
$\Lambda^p[w]=\left((\wedge^p TM)\otimes {\cal V}\right)[w]$.
Finally, we also introduce the following notation
$E_p[w]=\left((\wedge^p T^*M)\otimes \End({\cal V})\right)[w]$
for the bundle of endomorphism-valued densities of weight $w$
of type $(0,p)$ 
and
$E^p[w]=\big((\wedge^p TM)\otimes$ $\End({\cal V})\big)[w]$
for the bundle of endomorphism-valued densities of weight $w$
of type $(p,0)$.

The group $G$
is also promoted to a smooth vector bundle ${\cal G}[0]$
(gauge group), a subbundle of the bundle
$\End({\cal V})[0]$. 
Under the action of the gauge group ${\cal G}[0]$ 
the sections
$\varphi$ of the bundles ${\cal V}[w]$ and 
${\cal V}^*[w]$ transform as
\be
\varphi'(x)=U(x)\varphi(x)\,,\qquad
\bar\varphi'(x)=\bar\varphi(x)U^{-1}(x)\,,
\label{gt0}
\ee

The metric $E$ is assumed to be 
a section of the bundle
$({\cal V}^*\otimes{\cal V}^*)\left[\alpha\right]$
with some weight $\alpha$. 
It is invariant under
the gauge transformations
\be
E'(x)=(U^\dagger(x))^{-1}E(x)U^{-1}(x)=E(x)\,,
\ee
which guarantees the invariance of the fiber inner product.

However, to get a diffeomorphism-invariant $L^2$ inner product
\be
(\psi,\varphi)=\int\limits_M dx\,\left<\psi(x),\varphi(x)\right>\,,
\ee
and the $L^2$ norm
\be
||\varphi||^2=(\varphi,\varphi)=
\int\limits_M dx\,\left<\varphi(x),\varphi(x)\right>\,,
\ee
on the
vector bundle ${\cal V}[w]$ with the standard Lebesgue measure
$dx=dx^1\wedge\cdots\wedge dx^n$, the metric must be a {\it density}
of weight $(1-2w)$, more precisely, a section of the bundle 
$({\cal V}^*\otimes{\cal V}^*)[1-2w]$. 
Then
the completion of $C^\infty({\cal V}[w])$ in this norm defines
the Hilbert space $L^2({\cal V}[w])$.
Notice that this means, that
on the bundle ${\cal V}[{1\over 2}]$ of densities of 
weight $w={1\over 2}$ the metric $E(x)$ is invariant under the 
diffeomorphisms as well, and, therefore, 
without loss of generality can be assumed to
be {\it constant}.

\subsection{Noncommutative Metric}

Our goal is to construct
covariant self-adjoint second order differential operators acting
on a smooth sections of the bundles $\Lambda_p[{1\over 2}]$
and $\Lambda^p[{1\over 2}]$,
that are covariant under both diffeomorphisms,
\be
L'\varphi'(x')=J^{-1/2}(x)L\varphi(x)\,,
\ee
and the gauge transformations
\be
L'\varphi'(x)=U(x)L\varphi(x)\,.
\ee

To do this we need 
to define the following objects.
First of all, we need 
an isomorphism between the bundles 
$\Lambda_p[{1\over 2}]$ and $\Lambda^{p}[{1\over 2}]$.
This is usually achieved by the Hodge star operator, 
which is defined with the help of a Riemannian metric. This
is exactly the place where we want to {\it generalize the standard
theory} since we do not want to introduce a Riemannian
metric; instead of a Riemannian metric (which is an isomorphism
between the tangent $TM$ and the cotangent $T^*M$ bundles)
we introduce an {\it isomorphism} between the bundles
$\Lambda_1[w]=(T^*M\otimes{\cal V})[w]$ and 
$\Lambda^1[w]=(TM\otimes{\cal V})[w]$, i.e.
\be
a: \Lambda_1[w]\to 
\Lambda^1[w]\,.
\label{16}
\ee
Such an isomorphism is determined by a 
section of the vector bundle
$(TM\otimes TM\otimes \End({\cal V}))[0]$
defined 
by the matrix-valued symmetric tensor
$a^{\mu\nu}=\left(a^{\mu\nu}{}^A{}_B(x)\right)$
that satisfies the following conditions:
\begin{enumerate}
\item
The matrix $a^{\mu\nu}$ is self-adjoint 
(recall that $\overline{a^{\mu\nu}}=E^{-1}(a^{\mu\nu})^{\dagger}E$),
\be
\overline{a{}^{\mu\nu}}=a^{\mu\nu}\,
\ee
and symmetric,
\be
a^{\nu\mu}=a^{\nu\mu}\,.
\ee
\item
It transforms under the diffeomorphisms as
\be
a'^{\mu\nu}(x')={\partial x'^{\mu}\over\partial x^\alpha} {\partial
x'^{\nu}\over\partial x^\beta}a^{\alpha\beta}(x)\,,
\ee
and under the gauge transformations as
\be
a'^{\mu\nu}(x)=U(x) a^{\mu\nu}(x)U^{-1}(x)\,.
\ee

\item
Consider the matrix 
\be
H(x,\xi)=a^{\mu\nu}(x)\xi_\mu\xi_\nu\,,
\label{24}
\ee
with $\xi$
being a cotangent vector $\xi$ at a spacetime point $x$.  Since this
matrix is self-adjoint, all  its eigenvalues $h_{(i)}(x,\xi)$, 
$(i=1,\dots,s)$, must be real.
We will assume that they have constant multiplicities $d_i$.
The
eigenvalues are invariant under the gauge transformations (\ref{gt0})
and transform under the diffeomorphisms as 
\be
h'_i(x',\xi)=h_i(x,\xi')\,,
\ee
where 
\be
\xi'_{\mu}={\partial x^\alpha\over\partial x'^\mu} \xi_\alpha\,.
\ee
We will also require that this matrix satisfies one of the following
conditions. 
{\it Elliptic case}: The matrix  $H(x,\xi)$ is positive
definite, i.e. all  its eigenvalues are positive,  $h_{(i)}(x,\xi)>0$, for
any $x$ and any $\xi\ne 0$. 
{\it Hyperbolic case:} 
There is a one-form $\tau=\tau(x)$ (specifying the time direction)
such that at each point $x$ for each eigenvalue
$h_{(i)}(x,\tau(x))<0$, and for any cotangent vector $\zeta\ne 0$ not parallel
to $\tau$ each equation $h_{(i)}(x,\zeta+\lambda \tau(x))=0$,  has
exactly  two real distinct roots, $\lambda=\lambda_{\pm}(x,\zeta)$.  
More precisely, a second-order partial differential operator $L$
with the principal symbol $H(x,\xi)$ is {\it hyperbolic} 
if the roots $\lambda_j(x,\zeta)$ of the characteristic equation
\be
\det H(x,\zeta+\lambda\tau(x))=0
\ee
are real and {\it strictly hyperbolic} if they are real and distinct
\cite{egorov98,maslov76}).
\end{enumerate}

The infinitesimal forms of the diffeomorphisms and the gauge transformations
are obtained as follows.
Let $x'=f_t(x)$ be a one-parameter subgroup of diffeomorphisms
such that 
\be
f_t(x)|_{t=0}=x, \qquad {\rm and} \qquad
\xi(x)={d\over dt} f_t(x)\Big|_{t=0}\,.
\ee
Then
\be
\delta_\xi a^{\mu\nu}\equiv
{d\over dt} a'^{\mu\nu}(x')\Big|_{t=0}
=\xi^\alpha\partial_\alpha a^{\mu\nu}
-(\partial_\alpha\xi^\mu)a^{\alpha\nu}
-(\partial_\alpha\xi^\nu)a^{\mu\alpha}\,,
\ee

Similarly, let $U_t$ be a 
one-parameter subgroup of the gauge group
such that 
\be
U_t|_{t=0}=\II, \quad {\rm and} \qquad 
\omega={d\over dt} U_t\Big|_{t=0}. 
\ee
Then
\be
\delta_\omega a^{\mu\nu}
\equiv {d\over dt}a'^{\mu\nu}\Big|_{t=0}=[\omega,a^{\mu\nu}]\,.
\ee

Notice that these infinitesimal transformations are different from the gauge 
transformations studied in \cite{cutler87,wald87,boulanger00}
since there is only one vector field $\xi$.

%
%
%
%
%
%

\subsection{Finsler Geometry}

The above construction is closely related to Finsler geometry 
\cite{rund59}. First of all, we note that the eigenvalues $h_{(i)}(x,\xi)$
are homogeneous functions of $\xi$ of degree $2$, i.e.
\be
h_{(i)}(x,\lambda\xi)=\lambda^2 h_{(i)}(x,\xi)\,.
\ee
Next, we define the {\it Finsler metrics}
\be
g^{\mu\nu}_{(i)}(x,\xi)={1\over 2}
{\partial^2 \over\partial\xi_\mu\partial\xi_\nu }h_{(i)}(x,\xi)\,.
\ee
All these metrics are non-degenerate.
In the elliptic case all metrics $g_{(i)}^{\mu\nu}$ are positive definite;
in the hyperbolic case they have the signature $(-+\cdots+)$.
In the case when a Finsler metric does not depend on $\xi$
it is simply a Riemannian metric. 

The Finsler metrics are homogeneous functions of $\xi$ of degree $0$
\be
g_{(i)}^{\mu\nu}(x,\lambda\xi)=g_{(i)}^{\mu\nu}(x,\xi)\,,
\ee
so that they depend only on the direction of the
covector $\xi$ but not on its magnitude.
This leads to a number of identities, in particular,
\be
h_{(i)}(x,\xi)=g^{\mu\nu}_{(i)}(x,\xi)\xi_\mu\xi_\nu\,
\ee
and
\be
{1\over 2}{\partial\over\partial\xi_\mu} h_{(i)}(x,\xi)
=g^{\mu\nu}_{(i)}(x,\xi)\xi_\nu\,.
\ee

Next we define the inverse (covariant) Finsler metrics by
\be
g_{(i)}{}_{\mu\nu}(x,\dot x)g_{(i)}^{\nu\alpha}(x,\xi)=\delta^\alpha_\mu\,,
\ee
where $\dot x^\mu$ is the tangent vector defined by
\be
\dot x^\mu=g^{\mu\nu}_{(i)}(x,\xi)\xi_\nu\,,
\ee
so that
\be
\xi_\mu=g_{(i)}{}_{\mu\nu}(x,\dot x)\dot x^\nu\,.
\ee

Finally, this enables one to define the Finsler intervals
\be
ds^2_{(i)}=g_{(i)}{}_{\mu\nu}(x,\dot x)dx^\mu dx^\nu\,.
\label{33}
\ee
The existense of the Finsler metrics allows one to define various 
connections, curvatures etc (for details see \cite{rund59}).

As we see, the propagation of gauge fields induces Finsler geometry.

\subsection{Star Operators}

Since the map $a$ (\ref{16}) is an isomorphism, the 
inverse map 
\be
b=a^{-1}: \Lambda^1[w]\to 
\Lambda_1[w]\,.
\ee
is well defined. In other words,
for any $\psi^\mu=(\psi^{\mu A})$ there is a unique
$\varphi_\nu=(\varphi_\nu^A)$ satisfying the equation
$a^{\mu\nu}\varphi_\nu=\psi^\mu$, and, therefore,
there is a unique solution of the equations
\be
a^{\mu\nu}b_{\nu\alpha}=\delta^\mu_{\alpha}\,,\qquad
b_{\alpha\nu}a^{\nu\mu}=\delta^\mu_{\alpha}\,.
\ee
Notice that the matrix $b_{\mu\nu}$ has the property 
\be
\bar b_{\mu\nu} = b_{\nu\mu}\,,
\ee
but is neither symmetric $b_{\mu\nu}\ne b_{\nu\mu}$ nor self-adjoint
$\bar b_{\mu\nu}\ne b_{\mu\nu}$.

The isomorphism $a$ 
naturally defines the maps between
the bundles $\Lambda_p[w]$ and $\Lambda^{p}[w]$
\be
A: \Lambda_p[w]\to 
\Lambda^{p}[w]\,,\qquad
B: \Lambda^p[w]\to 
\Lambda_{p}[w]\,,
\ee
as follows
\be
(A\varphi)^{\mu_1\cdots\mu_p}
=A^{\mu_1\cdots\mu_p\nu_1\cdots\nu_p}
\varphi_{\nu_1\cdots\nu_p}\,,
\ee
where
\be
A^{\mu_1\cdots\mu_p\nu_1\cdots\nu_p}=
\Alt_{\mu_1\cdots\mu_p}\Alt_{\nu_1\cdots\nu_p} 
a^{\mu_1\nu_1}\cdots a^{\mu_p\nu_p}
\ee
and
\be
(B\varphi)_{\mu_1\cdots\mu_p}
=B_{\mu_1\cdots\mu_p\nu_1\cdots\nu_p}
\varphi^{\nu_1\cdots\nu_p}\,,
\ee
where
\be
B_{\mu_1\cdots\mu_p\nu_1\cdots\nu_p}
=\Alt_{\mu_1\cdots\mu_p}\Alt_{\nu_1\cdots\nu_p}
b_{\mu_1\nu_1}\cdots b_{\mu_p\nu_p}
\ee
Here $\Alt_{\mu_1\cdots\mu_p}$ 
denotes the complete antisymmetrization over
the indices $\mu_1,\dots,\mu_p$.

We will assume that these maps are isomorphisms as well. Strictly
speaking, one has to prove this.
This is certainly true for the weakly deformed maps
(maps close to the identity; more on this later).
Then the inverse operator
\be
A^{-1}: \Lambda^p[w]\to 
\Lambda_{p}[w]\,,
\ee
is defined by
\be
(A^{-1}\varphi)_{\mu_1\cdots\mu_p}
=A^{-1}_{\mu_1\cdots\mu_p\nu_1\cdots\nu_p}
\varphi^{\nu_1\cdots\nu_p}\,,
\ee
where $A^{-1}$ is determined by the equation
\be
A^{-1}_{\mu_1\cdots\mu_p\nu_1\cdots\nu_p}
A^{\nu_1\cdots\nu_p\alpha_1\cdots\alpha_p}
=\delta^{\alpha_1}_{[\mu_1}\cdots\delta^{\alpha_p}_{\mu_p]}\,.
\ee
Notice that because of the noncommutativity, the inverse operator 
$A^{-1}$ is not equal to the operator $B$, so that $A^{-1}B\ne\Id$.

This is used further to define the natural fiber inner product on the 
space of $p$-forms $\Lambda_p$ via
\be
\left<\psi,\varphi\right>
={1\over p!}\bar\psi_{\mu_1\cdots\mu_p}
A^{\mu_1\cdots\mu_p\nu_1\cdots\nu_p}
\varphi_{\nu_1\cdots\nu_p}\,.
\ee

We will also need a smooth self-adjoint
section $\rho$ (given locally by the matrix-valued function
$\rho=(\rho^A{}_B(x))$) of the bundle 
$\End({\cal V})[{1\over 2}]$ of endomorphism-valued densities of weight 
${1\over 2}$.
That is $\rho$ satisfies the equation
\be
\bar\rho=\rho\,,
\ee
(recall that $\bar\rho=E^{-1}\rho^\dagger E$), and transforms under 
diffeomorphisms as
\be
\rho'(x')=J^{-1/2}(x)\rho(x)\,,
\ee
and under the action of the gauge group ${\cal G}[0]$ as
\be
\rho'(x)=U(x)\rho(x) U^{-1}(x)\,.
\ee

Clearly, the matrices 
\be
a^{\mu\nu}=g^{\mu\nu}\II\,, \qquad
\rho=g^{1/4}\II\,,
\ee
where $g^{\mu\nu}$ is a 
(pseudo)-Riemannian metric and 
\be
g=|\det g^{\mu\nu}|^{-1}\,,
\ee
satisfy all the above conditions.
We will refer to this particular case as the {\it commutative limit}.

Of course, (on orientable manifolds)
we always have the standard volume form $\varepsilon$,
which is a section of the bundle $E_n[-1]$ given by
the completely antisymmetric Levi-Civita symbol 
$\varepsilon_{\mu_1\cdots\mu_n}$. 
The contravariant Levi-Civita
symbol $\tilde\varepsilon$ with components
\be
\varepsilon^{\mu_1\cdots\mu_n}=\sigma\varepsilon_{\mu_1\cdots\mu_n}\,,
\ee
with $\sigma=+1$ in the elliptic case and
$\sigma=-1$ in the hyperbolic case,
is a section of the bundle $E^n[1]$.

These densities are used to define the standard isomorphisms
between the densities bundles
\be
\varepsilon: \Lambda^p[w]\to \Lambda_{n-p}[w-1],\qquad
\tilde\varepsilon: \Lambda_p[w]\to \Lambda^{n-p}[w+1]
\ee
by
\be
(\varepsilon\varphi)_{\mu_1\cdots\mu_{n-p}}
={1\over p!}
\varepsilon_{\mu_1\cdots\mu_{n-p}\nu_1\cdots\nu_p}
\varphi^{\nu_1\cdots\nu_p}\,,
\quad
(\tilde\varepsilon\varphi)^{\mu_1\cdots\mu_{n-p}}
={1\over p!}
\varepsilon^{\mu_1\cdots\mu_{n-p}\nu_1\cdots\nu_p}
\varphi_{\nu_1\cdots\nu_p}\,.
\ee
By using the well known identity
\be
\varepsilon_{\mu_1\cdots\mu_{n-p}\nu_1\cdots\nu_p}
\varepsilon^{\mu_1\cdots\mu_{n-p}\lambda_1\cdots\lambda_p}
=\sigma (n-p)!p!
\delta^{\lambda_1}_{[\nu_1}\cdots\delta^{\lambda_p}_{\nu_p]}
\label{40}
\ee
we get
\be
\tilde\varepsilon\varepsilon
=\varepsilon\tilde\varepsilon
=\sigma (-1)^{p(n-p)}\Id\,.
\ee

By combining $\varepsilon$ and $\tilde\varepsilon$
with the endomorphism $\rho$ we can get the invariant form
$\varepsilon\rho^2$, which is a section of the bundle $E_n[0]$,
and the form $\tilde\varepsilon\rho^{-2}$, 
a section of the bundle $E^n[0]$.
Notice, however, that, in general, the contravariant form
$\tilde\varepsilon\rho^{-2}$ is not equal to that obtained
by raising indices of the covariant form $\varepsilon\rho^2$, i.e.
$\tilde\varepsilon\rho^{-2}\ne A\varepsilon\rho^2$ or
\be
\varepsilon^{\mu_1\cdots\mu_n}\rho^{-2}
\ne A^{\mu_1\cdots\mu_n\nu_1\cdots\nu_n}
\varepsilon_{\nu_1\cdots\nu_n}\rho^2\,.
\ee
If we require this to be the case then the matrix $\rho$ 
should be defined by
\be
\rho=\eta^{-1/4}\,,
\ee
where
\be
\eta=\sigma{1\over n!}\varepsilon_{\mu_1\cdots\mu_n}
\varepsilon_{\nu_1\cdots\nu_n}a^{\mu_1\nu_1}\cdots a^{\mu_n\nu_n}\,.
\ee
Clearly, $\eta$ is a density of weight $(-2)$. Since $a^{\mu\nu}$ is 
self-adjoint, we also find that $\eta$ and, hence, $\rho$ is self-adjoint.
The problem is that in general $\eta$ is not positive definite.
Notice that in the commutative limit 
\be
\eta=\sigma\,\det g^{\mu\nu}=|\det g_{\mu\nu}|^{-1}\,,
\ee
which is strictly positive.

Therefore, we can finally define
two different star operators
\be
*, \tilde *: \Lambda_p[w]\to \Lambda_{n-p}[w]
\ee
by
\be
*=\varepsilon\rho A\rho\,,\qquad
\tilde *=\rho^{-1}A^{-1}\rho^{-1}\tilde\varepsilon
\ee
that is
\be
(*\varphi)_{\mu_1\cdots\mu_{n-p}}
={1\over p!}\varepsilon_{\mu_1\cdots\mu_{n-p}\nu_1\cdots\nu_p}
\rho A^{\nu_1\cdots\nu_p\alpha_1\cdots\alpha_p}\rho
\varphi_{\alpha_1\cdots\alpha_p}\,,
\ee
\be
(\tilde *\varphi)_{\mu_1\cdots\mu_{n-p}}
={1\over p!}\rho^{-1}
(A^{-1})_{\mu_1\cdots\mu_{n-p}\beta_1\cdots\beta_{n-p}}
\rho^{-1}
\varepsilon^{\beta_1\cdots\beta_{n-p}\alpha_1\cdots\alpha_p}
\varphi_{\alpha_1\cdots\alpha_p}\,.
\ee
The star operators are self-adjoint in the sense
\be
\left<\varphi,*\psi\right>=\left<*\varphi,\psi\right>\,,\qquad
\left<\varphi,\tilde *\psi\right>=\left<\tilde *\varphi,\psi\right>\,,
\ee
and satisfy the relation: for any $p$ form
\be
*\tilde *=\tilde * *=\sigma (-1)^{p(n-p)}\Id\,.
\ee

\subsection{Hilbert Spaces}

Now, let us consider the bundles of densities of weight ${1\over 2}$,
${\cal V}[{1\over 2}]$, and, more generally, $\Lambda_p[{1\over 2}]$.
If $dx=dx^1\wedge \cdots \wedge dx^n$ 
is the standard Lebesgue measure in a local chart on $M$, then
we define the diffeo\-mor\-phism\--invariant $L^2$-inner product
\be
(\psi,\varphi)=\int\limits_M dx\,\left<\psi(x),\varphi(x)\right>\,,
\ee
and the $L^2$ norm
\be
||\varphi||^2=(\varphi,\varphi)=
\int\limits_M dx\,\left<\varphi(x),\varphi(x)\right>\,,
\ee
The completion of $C^\infty(\Lambda_p[{1\over 2}])$ in this norm defines
the Hilbert space $L^2(\Lambda_p[{1\over 2}])$.

\section{Differential Operators}

\subsection{Noncommutative Exterior Calculus}

Next, we define the invariant differential operators on smooth sections
of the bundles $\Lambda_p[0]$ and $\Lambda^p[1]$.
The exterior derivative (the {\it gradient}) 
\be
d: C^\infty(\Lambda_p[0])\to C^\infty(\Lambda_{p+1}[0])\,
\ee
is defined by
\be
(d \varphi)_{\mu_1\cdots\mu_{p+1}}
=(p+1)\partial_{[\mu_1}\varphi_{\mu_2\cdots\mu_p]}\,,
\qquad \mathrm{if}\ p=0,\dots,n-1\,,
\ee
\be
d\varphi=0 \qquad \mathrm{if} \ p=n\,,
\ee
where the square brackets denote the complete antisymmetrization.
The coderivative (the {\it divergence})
\be
\tilde d: C^\infty(\Lambda^p[1])\to C^\infty(\Lambda^{p-1}[1])\,
\ee
is defined by
\be
\tilde d=\sigma(-1)^{np+1}\tilde\varepsilon d\varepsilon\,.
\label{div}
\ee
By using (\ref{40}) one can easily find
\be
(\tilde d \varphi)^{\mu_1\cdots\mu_{p-1}}
=\partial_\mu \varphi^{\mu\mu_1\cdots\mu_{p-1}}\,
\qquad \mathrm{if} \ p=1,\dots,n\,,
\ee
\be
\tilde d\varphi=0 \qquad \mathrm{if} \ p=0\,.
\ee
One can also show that these definitions are covariant
and satisfy the standard relations
\be
d^2=\tilde d^2=0\,.
\ee

The endomorphism $\rho$ is a section of the bundle 
$\End({\cal V})[{1\over 2}]$.
Therefore, 
if $\varphi$ is a section of the bundle 
$\Lambda_p[{1\over 2}]$, the quantity
$\rho^{-1}\varphi$ is a section of the bundle 
$\Lambda_p[0]$. Hence, the derivative
$d(\rho^{-1}\varphi)$ is well defined as a smooth section of the
vector bundle $\Lambda_{p+1}[0]$.
By scaling back with the factor $\rho$ we get an invariant
differential operator
\be
\rho d\rho^{-1}: \ C^\infty\left(\Lambda_p
\left[{\scriptstyle{1\over 2}}\right]\right)\to
C^\infty\left(\Lambda_{p+1}
\left[{\scriptstyle{1\over 2}}\right]\right)\,.
\ee
Similarly, we can define the invariant operator of
codifferentiation on densities of weight ${1\over 2}$
\be
\rho^{-1}\tilde d\rho: \ C^\infty\left(\Lambda^p
\left[{\scriptstyle{1\over 2}}\right]\right)\to
C^\infty\left(\Lambda^{p-1}
\left[{\scriptstyle{1\over 2}}\right]\right)\,.
\ee

Now, let ${\cal B}$ be a smooth anti-self-adjoint
section of the vector bundle 
$E_1[0]$, 
defined by the matrix-valued covector
${\cal B}_\mu=({\cal B}_\mu{}^A{}_B(x))$ 
i.e.
\be
\bar{\cal B}_\mu=-{\cal B}_\mu\,,
\ee
that transforms under diffeomorphisms
as
\be
{\cal B}'_\mu(x')={\partial x^{\alpha}\over\partial x'^\mu}
{\cal B}_{\alpha}(x)\,,\qquad
\ee
and under the gauge transformations as
\be
{\cal B}'_\mu(x) = U(x){\cal B}_\mu(x) U^{-1}(x)
-(\partial_\mu U(x))U^{-1}(x)\,.
\ee
Such a section naturally defines the maps: 
\be
{\cal B}: \Lambda_p[w]\to \Lambda_{p+1}[w]
\ee
by
\be
({\cal B}\varphi)_{\mu_1\cdots\mu_{p+1}}=(p+1){\cal B}_{[\mu_1}
\varphi_{\mu_2\cdots\mu_{p+1}]}
\ee
and
\be
\tilde{\cal B}: \Lambda^p[w]\to\Lambda^{p-1}[w]
\ee
by
\be
(\tilde{\cal B}\varphi)^{\mu_1\cdots\mu_{p-1}}
={\cal B}_\mu\varphi^{\mu\mu_1\cdots\mu_{p-1}}\,.
\ee
Notice that
\be
\tilde{\cal B}=\sigma(-1)^{np+1}\tilde\varepsilon{\cal B}\varepsilon\,
\ee
similar to (\ref{div}).

This enables us to define the covariant
exterior derivative
\be
{\cal D}: C^\infty\left(\Lambda_p
\left[{\scriptstyle{1\over 2}}\right]\right)\to
C^\infty\left(\Lambda_{p+1}
\left[{\scriptstyle{1\over 2}}\right]\right)\,.
\ee
by
\be
{\cal D}=\rho(d+{\cal B})\rho^{-1}
\ee
and the covariant coderivative
\be
\tilde{\cal D}: C^\infty\left(\Lambda^p
\left[{\scriptstyle{1\over 2}}\right]\right)\to
C^\infty\left(\Lambda^{p-1}
\left[{\scriptstyle{1\over 2}}\right]\right)\,,
\ee 
by
\be
\tilde{\cal D}=\sigma(-1)^{np+1}\tilde\varepsilon
{\cal D}\varepsilon
=\rho^{-1}(\tilde d+\tilde{\cal B})\rho
\,.
\ee
These operators transform covariantly under both the
diffeomorphisms and the gauge 
transformations.

\subsection{Noncommutative Gauge Curvature}

One can easily show that 
the square of the operators ${\cal D}$ and $\bar{\cal D}$
\be
{\cal D}^2: 
C^\infty(\Lambda_p\left[{\scriptstyle{1\over 2}}\right])
\to C^\infty(\Lambda_{p+2}\left[{\scriptstyle{1\over 2}}\right])
\ee
\be
\tilde{\cal D}^2: 
C^\infty(\Lambda^{p+2}\left[{\scriptstyle{1\over 2}}\right])
\to C^\infty(\Lambda^p\left[{\scriptstyle{1\over 2}}\right])
\ee
are zero-order
differential operators. In particular, in the case $p=0$ they define
the gauge curvature ${\cal F}$,
which is a section 
of the bundle $E_2[0]$,
by
\be
({\cal D}^2\varphi)_{\mu\nu}
=\rho{\cal F}_{\mu\nu}\rho^{-1}\varphi\,,\qquad
\tilde{\cal D}^2\varphi
=\rho^{-1}{\cal F}_{\mu\nu}\rho\varphi^{\nu\mu}\,,
\ee
where
\be
{\cal F}=d{\cal B}+[{\cal B},{\cal B}]\,,
\ee
and the brackets $[\, ,\, ]$ denote the Lie bracket of two 
matrix-valued $1$-forms, i.e.
\be
[A,B]_{\mu\nu}=A_\mu B_\nu-B_\nu A_\mu\,.
\ee

The gauge curvature is anti-self-adjoint
\be
\bar{\cal F}_{\mu\nu}=-{\cal F}_{\mu\nu}
\ee
and
transforms covariantly under diffeomorphisms
\be
{\cal F}'_{\mu\nu}(x')={\partial x^\alpha\over\partial x'^\mu}
{\partial x^\beta\over\partial x'^\nu}{\cal F}_{\alpha\beta}(x)
\,,
\ee
and under the gauge transformations
\be
{\cal F}'_{\mu\nu}(x)=U(x){\cal F}_{\mu\nu}(x)
U^{-1}(x)\,.
\ee

\subsection{Noncommutative Laplacians}

Finally, by using the objects introduced above we can define
second-order differential operators
that are covariant under both diffeomorphisms,
and the gauge transformations.
In order to do that 
we need first-order differential operators (divergences) 
\be
\bar{\cal D}: \ C^\infty\left(\Lambda_p
\left[{\scriptstyle{1\over 2}}\right]\right)\to
C^\infty\left(\Lambda_{p-1}
\left[{\scriptstyle{1\over 2}}\right]\right)\,,
\ee

First of all, by using the $L^2$ inner product on the bundle
$\Lambda_p[{1\over 2}]$ (recall that the metric $E$ is
constant in this case) we define the adjoint operator
$\bar{\cal D}$
by 
\be
\left<\varphi,{\cal D}\psi\right>
=\left<\bar{\cal D}\varphi,\psi\right>\,.
\ee
This gives
\be
\bar{\cal D}=-A^{-1}\tilde {\cal D}A
=-\sigma(-1)^{np+1}A^{-1}\tilde\varepsilon{\cal D}\varepsilon A
=-A^{-1}\rho^{-1}(\tilde d+\tilde{\cal B})\rho A\,,
\ee
which in local coordinates reads
\be
(\bar{\cal D}\varphi)_{\mu_1\cdots\mu_p}
=-(A^{-1})_{\mu_1\cdots\mu_p\nu_1\cdots\nu_p}
\rho^{-1}(\partial_\nu+{\cal B}_\nu)\rho
A^{\nu\nu_1\cdots\nu_p\alpha_1\cdots\alpha_{p+1}}
\varphi_{\alpha_1\cdots\alpha_{p+1}}\,,
\ee
The problem with this definition is that usually it is
difficult to find the matrix 
$(A^{-1})_{\mu_1\cdots\mu_p\nu_1\cdots\nu_p}$.

Then we define the second order operators
\be
\bar{\cal D}{\cal D}\,,{\cal D}\bar{\cal D}, \Delta:\ C^\infty\left(\Lambda_p
\left[{\scriptstyle{1\over 2}}\right]\right)\to
C^\infty\left(\Lambda_{p}
\left[{\scriptstyle{1\over 2}}\right]\right)\,,
\ee
where the ``noncommutative Laplacian'' is
\bea
\Delta&=&
-\bar{\cal D}{\cal D}-{\cal D}\bar{\cal D}
\nonumber\\
&=&A^{-1}\tilde{\cal D}A{\cal D}+{\cal D}A^{-1}\tilde{\cal D}A
\nonumber\\
&=&
A^{-1}\rho^{-1}(\tilde d+\tilde{\cal B})\rho
A\rho(d+{\cal B})\rho^{-1}
\nonumber\\
&&
+\rho(d+{\cal B})\rho^{-1}A^{-1}\rho^{-1}
(\tilde d+\tilde{\cal B})\rho A\,.
\eea
In local coordinates this reads
\bea
(\Delta\varphi)_{\mu_1\cdots\mu_p}
&=&
\Biggl\{(p+1)A^{-1}_{\mu_1\dots\mu_p\nu_1\dots\nu_p}
\rho^{-1}(\partial_\nu+{\cal B}_\nu)\rho
\nonumber\\[5pt]
&&
\times
A^{\nu\nu_1\dots\nu_p\alpha\alpha_1\dots\alpha_p}
\rho(\partial_{\alpha}+{\cal B}_{\alpha})\rho^{-1}
\nonumber\\[5pt]
&&
+\rho(\partial_{[\mu_1}+{\cal B}_{[\mu_1})\rho^{-1}
A^{-1}_{\mu_2\dots\mu_{p-1}]\nu_1\dots\nu_{p-1}}
\nonumber\\[5pt]
&&
\times
\rho^{-1}
(\partial_\nu+{\cal B}_\nu)\rho
A^{\nu\nu_1\dots\nu_{p-1}\alpha_1\dots\alpha_p}
\Biggr\}
\varphi_{\alpha_1\dots\alpha_p}
\eea

We could have also defined the coderivatives by
\be
\bar{\cal D}_1=-* {\cal D}*
\,,\qquad
\bar{\cal D}_2=-B\tilde{\cal D}A
\,,\qquad
\bar{\cal D}_3=-\tilde * {\cal D}*
\,,\qquad
\bar{\cal D}_4=-*{\cal D}\tilde *\,.
\ee
These operators have the advantage that 
$\bar{\cal D}_1$ is
polynomial in the matrix $a^{\mu\nu}$
and $\bar{\cal D}_2$ is polynomial in the matrices
$a^{\mu\nu}$ and $b_{\mu\nu}$.
However, the second order operators
$\bar{\cal D}_j{\cal D}$, ${\cal D}\bar{\cal D}_j$
and $\Delta_j=-\bar{\cal D}_j{\cal D}-{\cal D}\bar{\cal D}_j$,
$(j=1,2,3,4)$,
are not self-adjoint, in general.
In the commutative limit all these definitions coincide
with the standard de Rham Laplacian.

In the special case $p=0$ the ``noncommutative
Laplacian" $\Delta$ reads
\be
\Delta=\rho^{-1}(\tilde d+\tilde{\cal B})\rho
A\rho(d+{\cal B})\rho^{-1}\,,
\ee
which in local coordinates has the form
\be
\Delta=\rho^{-1}(\partial_\mu+{\cal B}_\mu)\rho
a^{\mu\nu}\rho(\partial_\nu+{\cal B}_\nu)\rho^{-1}\,.
\ee

\subsection{Noncommutative Dirac Operator}

It is worth mentioning another approach.
Suppose we are given a self-adjoint section $\Gamma$ of the bundle 
$E_1[0]$, given locally by matrix valued vector
$\Gamma^\mu=(\Gamma^\mu{}^A{}_B(x))$ satisfying
\be
\bar\Gamma^\mu=\Gamma^\mu\,,
\ee
where as usual
$\bar\Gamma^\mu=E^{-1}(\Gamma^\mu)^\dagger E$.
It transforms under the diffeomorphisms as
\be
\Gamma'^\mu(x')={\partial x'^\mu\over\partial x^\alpha}\Gamma^\alpha(x)
\ee
and under the gauge transformation as
\be
\Gamma'^\mu(x)=U(x)\Gamma^\mu(x)U^{-1}(x)\,.
\ee
%
%
%
%
%
Then the matrix
\be
a^{\mu\nu}={1\over 2}\left(\Gamma^\mu\Gamma^\nu
+\Gamma^\nu\Gamma^\mu\right)
\ee
is self-adjoint $\overline{a^{\mu\nu}}=a^{\mu\nu}$ and symmetric
$a^{\mu\nu}=a^{\nu\mu}$.
Moreover, suppose this matrix
satisfies all the conditions for the matrix $a$ listed above
in Sect. 2.3.
Most importantly, the matrix
\be
H(x,\xi)=a^{\mu\nu}(x)\xi_\mu\xi_\nu=[\Gamma(x,\xi)]^2,
\ee
where 
\be
\Gamma(x,\xi)=\Gamma^\mu(x)\xi_\mu\,,
\ee
is nondegenerate and satisfies the ellipticity or
hyperbolicity conditions formulated in Sect. 2.3.
(The advantage of this approach is that in the elliptic case
the matrix $H(x,\xi)$ is
automatically positive definite.)

Such vector naturally defines the map
\be
\Gamma:\ C^\infty(\Lambda^p[w])\to C^\infty(\Lambda^{p+1}[w])
\ee
by
\be
(\Gamma\varphi)^{\mu_1\dots\mu_{p+1}}
=(p+1)\Gamma^{[\mu_1}\varphi^{\mu_2\dots\mu_{p+1}]}
\ee
and the map
\be
\tilde\Gamma:\ C^\infty(\Lambda_p[w])\to C^\infty(\Lambda_{p-1}[w])
\ee
by
\be
(\tilde\Gamma\varphi)_{\mu_1\dots\mu_{p-1}}
=\Gamma^\mu\varphi_{\mu\mu_1\dots\mu_p}\,.
\ee

Therefore, we can define
first-order invariant differential operator
(``noncommutative Dirac operator'')
\be
D: C^\infty\left(\Lambda_p
\left[{\scriptstyle{1\over 2}}\right]\right)\to
C^\infty\left(\Lambda_{p}
\left[{\scriptstyle{1\over 2}}\right]\right)\
\ee
by
\be
D=\tilde\Gamma {\cal D}
=\tilde\Gamma\rho(d+{\cal B})\rho^{-1}\,,
\ee
which in local coordinates reads
\be
(D\varphi)_{\mu_1\dots\mu_p}
=(p+1)\Gamma^\mu
\rho(\partial_{[\mu}+{\cal B}_{[\mu})\rho^{-1}\varphi_{\mu_1\dots\mu_p]}\,.
\ee

The adjoint of this operator with respect to the $L^2$ inner product
is
\be
\bar D=-A^{-1}\tilde {\cal D}\Gamma A
=-A^{-1}\rho^{-1}(\tilde d+\tilde{\cal B})\rho\Gamma A\,,
\ee
which in local coordinates becomes
\be
(\bar D\varphi)_{\mu_1\dots\mu_p}
=-(p+1)A^{-1}_{\mu_1\dots\mu_p\nu_1\dots\nu_p}
\rho^{-1}(\partial_{\nu}+{\cal B}_{\nu})\rho
\Gamma^{[\nu}A^{\nu_1\dots\nu_p]\alpha_1\dots\alpha_p}
\varphi_{\alpha_1\dots\alpha_p}\,.
\ee

These can be used then to define the second order self-adjoint
differential operators (``noncommutative Laplacians'')
\bea
\Delta_1&=&-D \bar D
\nonumber\\
&=&\tilde\Gamma{\cal D}A^{-1}\tilde{\cal D}\Gamma A
\nonumber\\
&=&\tilde\Gamma\rho(d+{\cal B})\rho^{-1}A^{-1}\rho^{-1}
(\tilde d+\tilde{\cal B})\rho\Gamma A\,,
\eea
\bea
\Delta_2&=& -\bar D D
\nonumber\\
&=&A^{-1}\tilde{\cal D}\Gamma A \tilde\Gamma {\cal D}
\nonumber\\
&=&A^{-1}\rho^{-1}(\tilde d+\tilde{\cal B})\rho
\Gamma A\tilde\Gamma \rho(d+{\cal B})\rho^{-1}\,.
\eea

In the case $p=0$ these operators have the form
\bea
\Delta_1&=&\tilde\Gamma{\cal D}\tilde{\cal D}\Gamma
\nonumber\\
&=&\Gamma^\mu \rho(\partial_\mu+{\cal B}_\mu)
\rho^{-2}(\partial_\nu+{\cal B}_\nu)\rho\Gamma^\nu\,,
\\
\Delta_2&=&\tilde{\cal D}\Gamma\tilde\Gamma{\cal D}
\nonumber\\
&=&\rho^{-1}(\partial_\nu+{\cal B}_\nu)
\rho\Gamma^\nu \Gamma^\mu
\rho(\partial_\mu+{\cal B}_\mu)\rho^{-1}\,.
\eea

In the elliptic case the above constructions can be used to develop
noncommutative generalization of the standard
Hodge-de Rham theory, in particular, noncommutative versions of
the index theorems, the cohomology groups, the heat kernel etc.
This is a very interesting topic that requires further study.

\subsection{Differential Operators for $p=0$}

In present paper we will restrict ourselves to the case $p=0$,
more precisely, to the second order
self-adjoint differential operators acting on smooth sections
of the bundle ${\cal V}[{1\over 2}]$ that can be presented in the form
\be
L=-\rho^{-1}(\partial_\mu+{\cal B}_\mu)\rho
a^{\mu\nu}\rho(\partial_\nu+{\cal B}_\nu)\rho^{-1}+Q\,,
\label{do}
\ee
by an appropriate choice of
$a^{\mu\nu}$, $\rho$, ${\cal B}_\mu$ and 
$Q\in C^\infty(\End({\cal V})[0])$.
It is this operator that we will study in detail below.

This operator can be also written in the form
\be
L=\bar X_\mu a^{\mu\nu}X_\nu+Q\,,
\ee
where
\be
X_\mu=\partial_\mu+{\cal C}_\mu\,,
\label{129}
\ee
\be
\bar X_\mu=-\partial_\mu+\bar{\cal C}_\mu\,,
\ee
and
\be
{\cal C}_\mu=-\rho_{,\mu}\rho^{-1}+\rho{\cal B}_\mu\rho^{-1}\,,
\label{119}
\ee
\be
\bar{\cal C}_\mu=-\rho^{-1}\rho_{,\mu}-\rho^{-1}{\cal B}_\mu\rho\,,
\ee
comma denotes partial derivatives, e.g. $\rho_{,\mu}=\partial_\mu\rho$,

In more details, it can be presented in the form
\be
L=-a^{\mu\nu}\partial_\mu\partial_\nu+Y^\mu\partial_\mu+Z\,,
\ee
where
\bea
Y^\mu
&=&
-a^{\mu\nu}{}_{,\nu}
+\bar{\cal C}_\nu a^{\mu\nu}
-a^{\mu\nu}{\cal C}_\nu
\\[12pt]
Z&=&
Q-(a^{\mu\nu}{\cal C}_\nu)_{,\mu}+\bar{\cal C}_\mu a^{\mu\nu}{\cal C}_\nu
\eea


The leading symbol of the operator $L$
is given by the matrix $H(x,\xi)=a^{\mu\nu}(x)\xi_\mu\xi_\nu$,
with $\xi$ a cotangent vector.
In the Sect 2.3 we formulated precise conditions on this matrix
for the operator $L$ to be elliptic or hyperbolic.
The system of hyperbolic partial differential equations describes the
propagation of a {\it collection of waves} and 
generates the causal structure on the spacetime manifold
\cite{avramidi03a,avramidi03b}. 
In the following we restrict ourselves for simplicity to the
elliptic case, in which the leading symbol $H(x,\xi)$ is
positive definite.

%
%
%

\section{Examples}

We see that in the matrix case the operator $L$ does not define a unique
Riemannian metric $g^{\mu\nu}$. Rather there is a matrix-valued symmetric 2-tensor
field $a^{\mu\nu}$. 
Although the matrix $g^{\mu\nu}$ plays the role of the Riemannian
metric in the commutative limit, it loses this role in fully
noncommutative strongly deformed theory.

\begin{description}

\item[Abelian Case.] This is the case studied by Cutler and Wald
\cite{cutler87,wald87} and Boulanger et al \cite{boulanger00}.
In this case the matrix $a^{\mu\nu}$ has the following form
\be
a^{\mu\nu}=g^{\mu\nu}_{(a)}\Pi_{(a)}\,, \qquad 
\rho=g^{1/4}_{(a)}\Pi_{(a)}\,,
\ee
where $g_{(a)}=(\det g^{\mu\nu}_{(a)})^{-1}$, and 
$\Pi_{(a)}$ are the projections on the $a$-th component,
that is diagonal matrices such that $(\Pi_{(a)})^A{}_B=\delta^A_B
\delta^A_{(a)}\delta^{(a)}_B$ (no summation !).
In this case all the structures commute and there is really
nothing new in the matrix gravity theory, which is just the
sum of the Einsten-Hilbert actions. What we consider in the present 
paper is {\it radically different}
 since it involves {\it fully noncommutative 
theory}. Some of the examples are considered below.

\item[Commutative Limit.]
The simplest case is when there is a decomposition
\be
a^{\mu\nu}=g^{\mu\nu}\II+\varkappa h^{\mu\nu}\,,
\qquad
\rho=g^{1/4}\exp(\varkappa \phi)\,,
\label{41}
\ee
where $\II$ is the unity matrix,
$g^{\mu\nu}$ is a contravariant symmetric $2$-tensor, 
$h^{\mu\nu}$ is a matrix-valued tensor field,
$\phi$ is a matrix-valued scalar field and
$\varkappa$ is a deformation parameter.
%
%

\item[Reducible Case.]
A more general decomposition is given by
\be
a^{\mu\nu}=g^{\mu\nu}\Omega+\varkappa h^{\mu\nu}\,,
\qquad
\rho=g^{1/4}\Omega^{-n/8}\exp(\varkappa \phi)\Omega^{-n/8}\,,
\label{def0}
\ee
where $\Omega$ is a self-adjoint positive definite (non-constant) 
section of the endomorphism bundle $\End({\cal V})[0]$,
i.e. $\bar\Omega=\Omega$, $\Omega>0$.

\item[Commutative $N=2$ Case.]
The simplest nontrivial 
model is when the matrix $a^{\mu\nu}$ is a real symmetric
$2\times 2$ matrix with the gauge group $O(2)$ and the fiber metric 
$E=\II$. We have then the following decomposition
\be
a^{\mu\nu}=g^{\mu\nu}\II+\varkappa\tau h^{\mu\nu}\,,
\qquad
\rho=g^{1/4}\exp(\varkappa\tau\phi)\,,
\ee
where $h^{\mu\nu}$ is a tensor field, $\phi$ is a scalar and
\be
\tau=\left(\begin{array}{rr}
0 & 1\\
1 & 0\\
\end{array}
\right)\,.
\ee
The condition of positivity of the 
matrix $H(x,\xi)=a^{\mu\nu}\xi_\mu\xi_\nu$ 
reads
\be
\varkappa | h^{\mu\nu}\xi_\mu\xi_\nu|<
|g^{\mu\nu}\xi_\mu\xi_\nu|
\ee
for any $\xi\ne 0$. 

\item[Noncommutative $N=2$ Case.]
The simplest example of a fully deformed noncommutative 
theory is the case of $N=2$ with the gauge group $U(2)$. 
Assuming that the fiber metric is given by
\be
E=\II\,,
\ee
we have the decomposition in terms of Pauli matrices
\be
a^{\mu\nu}=g^{\mu\nu}\II+\varkappa\tau_a h^{\mu\nu}_a\,,
\qquad
\rho=g^{1/4}\exp(\varkappa\tau_a\phi_a)
\ee
where $h^{\mu\nu}_a$ are tensor fields, $\phi_a$ are scalars 
and $\tau_a$ are standard Pauli matrices
\be
\tau_1=\left(\begin{array}{rr}
0 & 1\\
1 & 0\\
\end{array}\right)\,,\qquad
\tau_2=\left(\begin{array}{rr}
0 & -i\\
i & 0\\
\end{array}\right)\,,\qquad
\tau_3=\left(\begin{array}{rr}
1 & 0\\
0 & -1\\
\end{array}\right)\,.
\ee
The condition of positivity of the 
matrix $H(x,\xi)=a^{\mu\nu}\xi_\mu\xi_\nu$ 
reads
\be
\left(g^{\mu\nu}g^{\alpha\beta}
-\varkappa^2 h^{\mu\nu}_a h^{\alpha\beta}_a
\right)
\xi_\mu\xi_\nu\xi_\alpha\xi_\beta>0
\ee
for any $\xi\ne 0$. 

\item[Large $N$ Case.]
It is also interesting to consider the case of the $U(N)$ gauge group
in the limit $N\to \infty$. Since the matrix-valued metrics do not 
commute, it could serve as a toy model 
for quantum gravity. The decomposition has the form
\be
a^{\mu\nu}=g^{\mu\nu}\II+\varkappa(N) h^{\mu\nu}_a T_a\,,
\ee
where $T_a$ are the generators of $SU(N)$ and one can adjust
the dependence of the coupling $\varkappa$ on $N$ to simplify
the limit $N\to\infty$.

\end{description}

\section{Spectral Asymptotics}

As we mentioned above we restrict ourselves to elliptic operators.
All the geometric quantities we need for the gravity
theory can be extracted
from the spectrum of the elliptic operator $L$ (\ref{do})
(on a compact manifold $M$). 
The spectrum of the operator $L$ is defined as usual by
\be
L\varphi_k=\lambda_k\varphi_k\,, 
\qquad ||\varphi_k||_{L^2}=1\,,
\ee
where $k=1,2,\dots$. It is well known that a self-adjoint elliptic
partial differential operator with positive definite leading symbol on a
compact manifold without boundary has a discrete real spectrum bounded
from below \cite{gilkey95}. 
Since the operator $L$ transforms covariantly under the
diffeomorphisms as well as under the gauge transformations (\ref{gt0})
the spectrum is {\it invariant} under these transformations.

To study the spectral asymptotics, i.e. the behavior of $\lambda_k$ as
$k\to \infty$, one introduces the spectral functions, in particular,
the heat trace
\be
\Theta(t)=\sum_{k=1}^\infty e^{-t\lambda_k}\,,
\label{hksp}
\ee
and the zeta function
\be
\zeta(s)=\mu^{2s}\sum_{k=1}^\infty(\lambda_k+m^2)^{-s}\,,
\label{zetasp}
\ee
where each eigenvalue is taken with its multiplicity and
$\mu$ is a renormalization parameter introduced to preserve dimensions.
Here $t$ is a real parameter, $m^2$ is a
sufficiently large positive mass parameter (so that the operator
$(L+m^2)$ is positive) and $s$ is a complex parameter.
It is well
known that for a self-adjoint elliptic second-order  partial
differential operator with positive definite leading symbol the series
(\ref{hksp}) converges and defines a smooth function for $t>0$
and the series (\ref{zetasp}) converges for $\mathrm{Re}\, s > n/2$ and
defines a meromorphic function with simple isolated poles on the
real line.
These spectral functions are related by the Laplace-Mellin transform
\be
\zeta(s)={\mu^{2s}\over\Gamma(s)}\int\limits_0^\infty dt\, t^{s-1}\,
e^{-tm^2}\Theta(t)\,.
\ee
Moreover, it is also known that $\zeta(s)$ is analytic at $s=0$, 
which enables one to define the
regularized determinant of the operator $(F+m^2)$
(the one-loop effective action in quantum theory)
\cite{dewitt84}
\be
\Gamma_{(1)}={1\over 2}\log\Det \left({L+m^2\over \mu^2}\right)
=-{1\over 2}{\partial_s}\zeta(s)\Big|_{s=0}\,.
\ee

It turns out that the study of the spectral asymptotics is equivalent to
the study of the asymptotic expansion of the heat trace $\Theta(t)$ as
$t\to 0$  and that of the zeta function and the effective action as
$m^2\to\infty$. 
It is well known that  there is an asymptotic expansion of the
heat trace invariant as $t\to 0^+$ \cite{gilkey95}
\be
\Theta(t)\sim (4\pi)^{-n/2}
\sum\limits_{k=0}^\infty t^{k-{n\over 2}} A_{k}\,,
\label{17}
\ee
and an asymptotic expansion of the zeta-function as $m\to\infty$
\be
\zeta(s)\sim (4\pi)^{-n/2}
\sum\limits_{k=0}^\infty {\Gamma(k+s-{n\over 2})\over
\Gamma(s)}\mu^{2s}m^{n-2s-2k}
A_{k}\,,
\label{18}
\ee
where $A_{k}$ are so-called heat invariants 
\cite{gilkey95,avramidi00,avramidi91b,avramidi99,avramidi98}.
The coefficients $A_{k}$ are spectral invariants  determined by the
integrals over the manifold of some local invariants
\cite{gilkey95,avramidi91b,avramidi00} (for a review, see
\cite{avramidi99,avramidi02}) constructed from the coefficients of the
operator $L$ and their derivatives so that  they are polynomial in
the derivatives of the  coefficients of the operator $L$.
The heat invariants $A_k$ determine further
the large mass asymptotic expansion of the effective action as
$m\to\infty$ \cite{avramidi91b,avramidi99,avramidi00,avramidi02}
\be
\Gamma_{(1)}\sim c_0m^nA_0+c_1m^{n-1}A_1
+\sum\limits_{k=2}^\infty c_km^{n-2k}A_k\,,
\label{153}
\ee
where
\be
c_k=-{1\over 2}(4\pi)^{-n/2}{\partial\over\partial s}\left.\left[
{\Gamma(k+s-{n\over 2})\over\Gamma(s)}\left(m\over \mu\right)^{-2s}
\right]\right|_{s=0}
\ee

One of the most powerful methods to study the spectral asymptotics is the 
heat kernel.
The heat kernel, $U(t|x,x')$, is defined as the
kernel of the heat semigroup $\exp(-tL)$ operator for $t>0$, i.e.  the
fundamental solution of the heat equation
\be
(\partial_t+L)U(t|x,x')=0
\ee
with the initial condition
\be
U(0^+|x,x')=\delta(x-x')\,,
\ee
$\delta(x-x')$ being the Dirac distribution. 

{}For $t>0$ the heat kernel
$U(t|x,x')$ is a smooth function near the diagonal of $M\times M$ and has a
well defined diagonal value
\be
U(t|x,x)=\sum_{k=1}^\infty 
e^{-t\lambda_k}\varphi_k(x)\otimes\bar\varphi_k(x)\,,
\ee
which has the asymptotic
expansion as $t\to 0$ \cite{gilkey95}
\be
U(t|x,x)\sim (4\pi)^{-n/2}
\sum_{k=0}^\infty t^{k-{n\over 2}} a_k(x)\,,
\ee
where $a_k(x)$ are some covariant densities.

Moreover, the heat semigroup $\exp(-tL)$ is a 
trace-class operator with a well defined $L^2$ trace
\be
\Tr_{L^2}\exp(-tL)
=\int\limits_M \,dx\,\tr_V U(t|x,x)
=\Theta(t)\,,
\ee
where $\tr_V$ denotes the fiber trace. 
We have defined the heat kernel in such
a way that it transforms as a density of weight ${1\over 2}$ at both points
$x$ and $x'$. More precisely, it is a section of the exterior tensor
product bundle ${\cal V}[{1\over 2}]\boxtimes{\cal V}^*[{1\over 2}]$.
Therefore, the heat kernel diagonal transforms as a density of
weight $1$, i.e. it is a section of the bundle
$\End({\cal V})[1]$, and the trace $\Tr_{L^2}\exp(-tL)$ is invariant under
diffeomorphisms.

Therefore, the heat invariants $A_k$ can be constructed by
computing the $t\to 0$ asymptotics of the solution of  the heat
equation and integrating the coefficients $a_k(x)$ over the manifold,
i.e.
\be
A_k=\int\limits_M dx\, \tr_V a_k(x)\,.
\ee

A second-order differential operator is called Laplace type
if it has a scalar leading symbol.
Most of the calculations in quantum
field theory and spectral geometry
are restricted to the Laplace type operators for
which nice theory of heat kernel asymptotics is available
\cite{gilkey95,avramidi91b,avramidi98,avramidi99,
avramidi00,avramidi02}.
However, the operators condidered in the present paper have a matrix
valued principal symbol $H(x,\xi)$ and are, therefore, not of
Laplace type. The study of heat kernel asymptotics
for non-Laplace type operators is quite new and the methodology is still
underdeveloped. As a result even the first heat invariants
($A_0$, $A_1$ and $A_2$) are not known in general. For some partial
results see 
\cite{gilkey91,avrbrans01,avrbrans02}.

\subsection{Calculation of the Invariants $A_0$ and $A_1$}

For so called natural non-Laplace type differential 
operators, which are constructed from
a Riemannian metric and canonical connections on spin-tensor bundles
the coefficients $A_0$ and $A_1$ were computed in \cite{avrbrans02}.
Following this paper we will use a {\it formal} method that is sufficient
for our purposes of computing the asymptotics of the heat trace
of the second-order elliptic self-adjoint operator $L$ (\ref{do}).

First, we present the heat kernel diagonal in the form
\bea
U(t|x,x)
&=&\int\limits_{\RR^n} {d\xi\over (2\pi)^n} e^{-i\xi x}\exp(-tL) e^{i\xi x} 
\,,
\eea
where $\xi x=\xi_\mu x^\mu$, which can be transformed to
\bea
U(t|x,x)&=&\int\limits_{\RR^n} {d\xi\over (2\pi)^n} 
\exp\left[-t\left(H+K+L\right)\right]\cdot \II\,,
\eea
where $H$ is the leading symbol of the operator $L$
\be
H=a^{\mu\nu}\xi_\mu\xi_\nu\,,
\ee
and $K$ is a first-order self-adjoint operator defined by
\be
K=i\xi_\mu\left(\bar X_\nu a^{\mu\nu}-a^{\mu\nu}X_\nu\right)\,,
\ee
where the operator $X_\mu$ is defined by (\ref{129}).
Here the operators in the exponent act on the unity matrix $\II$
from the left.
By changing the integration variable $\xi\to t^{-1/2}\xi$ we obtain
\bea
U(t|x,x)&=&(4\pi t)^{-n/2}\int\limits_{\RR^n} {d\xi\over \pi^{n/2}} 
\exp\left(-H-\sqrt tK-tL\right)\cdot \II\,,
\eea
and the problem becomes now to evaluate the first three terms
of the asymptotic expansion of this integral as $t\to 0$.
By using the Volterra series
\bea
\exp(A+B)&=&e^A+\sum\limits_{k=1}^\infty
\int\limits_0^1 d\tau_k \int\limits_0^{\tau_k}d\tau_{k-1}\cdots 
\int\limits_0^{\tau_2} d\tau_1
\nonumber\\[10pt]
&&\times
e^{(1-\tau_{k})A} Be^{(\tau_k-\tau_{k-1}) A}
\cdots
e^{(\tau_2-\tau_1)A} Be^{\tau_1 A}\,,
\eea
we get
\bea
&&\exp\left(-H-\sqrt tK-tL\right)
=e^{-H}
\nonumber\\
&&
-t^{1/2}\int\limits_0^1 d\tau_1 e^{-(1-\tau_1)H}K e^{-\tau_1 H}
\nonumber\\
&&
+t\Biggl[
\int\limits_0^1d\tau_2\int\limits_0^{\tau_2}d\tau_1 e^{-(1-\tau_2)H}
K e^{-(\tau_2-\tau_1)H}Ke^{-\tau_1 H}
-\int\limits_0^1 d\tau_1 e^{-(1-\tau_1)H}Le^{-\tau_1 H}
\Biggr]
\nonumber\\[10pt]
&&
+O(t^2)
\,.
\eea
Now, since $K$ is linear in $\xi$ the term proportional to $t^{1/2}$
vanishes after integration over $\xi$. Thus, we obtain
the first two coefficients of the heat kernel diagonal
\be
U(t|x,x)\sim (4\pi t)^{-n/2}\left[a_0+ta_1+O(t^2)\right]
\ee
in the form
\bea
a_0 &=&\int\limits_{\RR^n}{d\xi\over \pi^{n/2}}\,
e^{-H}\,,
\\
a_1&=&\int\limits_{\RR^n}{d\xi\over \pi^{n/2}}\,
\Biggl[
\int\limits_0^1d\tau_2\int\limits_0^{\tau_2}d\tau_1 e^{-(1-\tau_2)H}
K e^{-(\tau_2-\tau_1)H}Ke^{-\tau_1 H}
\nonumber\\
&&
-\int\limits_0^1 d\tau_1 e^{-(1-\tau_1)H}Le^{-\tau_1 H}
\Biggr] 
\,.
\eea
These quantities are matrix densities that define finally the heat
invariants
\bea
A_0 &=&\int\limits_M dx\,\int\limits_{\RR^n}{d\xi\over \pi^{n/2}}\,
\tr_V\,e^{-H}\,,
\\
A_1&=&\int\limits_M dx\,\int\limits_{\RR^n}{d\xi\over \pi^{n/2}}\,
\tr_V\,\Biggl[
\int\limits_0^1d\tau_2\int\limits_0^{\tau_2}d\tau_1 e^{-(1-\tau_2)H}
K e^{-(\tau_2-\tau_1)H}Ke^{-\tau_1 H}
\nonumber\\
&&
-\int\limits_0^1 d\tau_1 e^{-(1-\tau_1)H}Le^{-\tau_1 H}
\Biggr] 
\,.
\eea
These quantities are invariant under both the diffeomorphisms
and the gauge transformations. The coefficient $a_0$ is constructed 
from the matrix $a$ but not its derivatives, whereas the coefficient
$a_1$ is constructed from the matrix $a$ and its first and second 
derivatives as well as from the first derivatives of the field ${\cal B}$
and the matrix $\rho$ and its first and second derivatives;
obviously, it is linear in the matrix $Q$.
Morevover, it does not depend on the 
derivatives of $Q$ and is polynomial in
the derivatives of $a^{\mu\nu}$, $\rho$ and ${\cal B}_\mu$,
more precisely, linear in second derivatives of $a$ and $\rho$
and the first derivatives of ${\cal B}$ and quadratic in 
first derivatives of $a$ and $\rho$.
Further, since the operator $L$ is self-adjoint, the heat kernel
diagonal is self-adjoint matrix density and the heat trace is
a real invariant. Therefore, the coefficients $a_0$ and $a_1$ are
self-adjoint, and the invariants $A_0$ and $A_1$ are real.

By integrating by parts and using the cyclic property of the trace
one can simplify the expression for the invariant
$A_1$ as follows
\bea
A_1&=&\int\limits_M dx\,\int\limits_{\RR^n}{d\xi\over \pi^{n/2}}\,
\tr_V\,\Biggl[
-e^{-H}Q
-\int\limits_0^1 d\tau_1 
\overline{W_\mu(1-\tau_1)}a^{\mu\nu}W_\nu(\tau_1)
\nonumber\\
&&
+\int\limits_0^1d\tau_2\int\limits_0^{\tau_2}d\tau_1 
\overline{T(1-\tau_2)} 
e^{-(\tau_2-\tau_1)H}
T(\tau_1)
\Biggr] 
\,,
\eea
where
\bea
W_\mu(\tau)&=&X_\mu e^{-\tau H}\,,\\[10pt]
T(\tau)&=&K e^{-\tau H}\,.
\eea

Finally, by using the Duhammel formula
\be
\partial A=\int\limits_0^1 ds\, e^{(1-s)A}(\partial A)e^{s A}
\ee
we can compute the derivatives of the exponential
$e^{-\tau H}$
\bea
\partial_\mu e^{-\tau H}&=&
-\int\limits_0^\tau ds\, e^{-(\tau-s)H}H_\mu
e^{-sH}\,,
\eea
where
\be
H_\mu=a^{\alpha\beta}{}_{,\mu}\xi_\alpha\xi_\beta\,.
\label{162}
\ee
We can use this formula to compute the second derivatives of 
$e^{-\tau H}$ needed for the local coefficient $a_1$
\bea
\partial_\mu\partial_\nu e^{-\tau H}
&=&
-\int\limits_0^\tau ds_1\, e^{-(\tau-s_1)H}H_{\mu\nu}e^{-s_1 H}
\nonumber\\
&&
+\int\limits_0^\tau ds_2\int\limits_0^{s_2} ds_1\Biggl[
e^{-(s_2-s_1)H}H_\nu e^{-s_1H}H_\mu e^{-(\tau-s_2)H}
\nonumber\\
&&
+e^{-(\tau-s_2)H}H_\mu e^{-(s_2-s_1)H}H_\nu e^{-s_1H}
\Biggr]\,,
\eea
where
\be
H_{\mu\nu}=a^{\alpha\beta}{}_{,\mu\nu}\xi_\alpha\xi_\beta\,.
\ee

By isolating the overall exponential factor
we get
\bea
\partial_\mu e^{-\tau H}&=&-\beta_\mu(\tau) e^{-\tau H}
\,,
\eea
where
\be
\beta_\mu(\tau)=\int\limits_0^\tau ds\, e^{-sH}H_\mu e^{sH}\,,
\label{166}
\ee
which can be presented in the algebraic form
\be
\beta_\mu(\tau)={e^{-\tau Ad_H}-1\over Ad_H}H_\mu
=\sum_{k=1}^\infty {(-\tau)^k\over k!}
\underbrace{[H,\cdots [H}_{k-1},H_\mu]\cdots]\,.
\ee
Here $Ad_H$ is the operator defined by 
\be
Ad_H A=[H,A]\,.
\ee

By using the above formulas we obtain
\be
W_\mu(\tau)=[{\cal C}_\mu-\beta_\mu(\tau)]e^{-\tau H}
\,,
\ee
\be
T(\tau)=i[\alpha+2J^\nu\beta_\nu(\tau)]e^{-\tau H}\,,
\ee
where
\bea
J^\nu &=& a^{\mu\nu}\xi_\mu\,,
\label{171}\\
\alpha &=& \bar{\cal C}_\nu J^\nu-J^\nu{\cal C}_\nu
-\xi_\mu a^{\mu\nu}{}_{,\nu}\,.
\label{172}
\eea

Thus, finally, summarizing this section we 
formulate the results in form of a 
theorem.

\begin{theorem}
The heat trace of the operator $L$ (\ref{do}) has the asymptotic expansion
as $t\to 0^+$
\be
{\rm Tr}_{L^2}\exp(-tL)= 
(4\pi t)^{-n/2}\left[A_0+tA_1+O(t^{2})\right]\,,
\label{t1}
\ee
where
\bea
A_0 &=&\int\limits_M dx\,\int\limits_{\RR^n}{d\xi\over \pi^{n/2}}\,
\tr_V\,e^{-H}\,,
\\
A_1&=&\int\limits_M dx\,\int\limits_{\RR^n}{d\xi\over \pi^{n/2}}\,
\tr_V\,e^{-H}
\Biggl\{-Q
\nonumber\\
&&
-\int\limits_0^1 d\tau_1 
[\bar{\cal C}_\mu-\bar\beta_\mu(1-\tau_1)]
a^{\mu\nu}[{\cal C}_\nu-\beta_\nu(\tau_1)]
\nonumber\\
&&
+\int\limits_0^1d\tau_2\int\limits_0^{\tau_2}d\tau_1 
e^{(\tau_2-\tau_1)H}
[\bar\alpha+2\bar\beta_\nu(1-\tau_2)J^\nu]
e^{-(\tau_2-\tau_1)H}
\nonumber\\
&&\times
[\alpha+2J^\mu\beta_\mu(\tau_1)]
\Biggr\}
\,.
\eea
Here ${\cal C}_\mu$, $\beta_\mu$, $\alpha$ and $J^\nu$ are given by
eqs. (\ref{119}), (\ref{162}), (\ref{166}), 
(\ref{171}) and (\ref{172}).

\end{theorem}


\subsection{Commutative Limit}

Let us compute the coefficient $A_1$ in the commutative limit.
We let
\be
a^{\mu\nu}=g^{\mu\nu}\II\,,\qquad
\rho=g^{1/4}e^{\phi}\II\,,\qquad
Q=q\II\,,
\label{547}
\ee
where $g^{\mu\nu}$ is a nonsingular matrix,
$
g=(\det g^{\mu\nu})^{-1}\,,
$
and $\phi$
is a scalar function.
The operator $L$ (\ref{do}) has the form
\bea
L&=&-g^{-1/4}(\partial_\mu+{\cal B}_\mu
+\phi_{,\mu})g^{1/2}
g^{\mu\nu}(\partial_\nu+{\cal B}_\nu
-\phi_{,\nu})g^{-1/4}+q
\nonumber\\
&=&-g^{-1/4}(\partial_\mu+{\cal B}_\mu)g^{1/2}
g^{\mu\nu}(\partial_\nu+{\cal B}_\nu)g^{-1/4}+\tilde q\,,
\eea
where
\be
\tilde q=q+\Box\phi+g^{\mu\nu}\phi_{,\mu}\phi_{,\nu}\,.
\label{194}
\ee
and 
\be
\Box=g^{-1/2}\partial_\mu g^{1/2}g^{\mu\nu}\partial_\nu\,.
\ee

Then
\be
H=g^{\mu\nu}\xi_\mu\xi_\nu\,,\qquad 
H_\mu = g^{\alpha\beta}{}_{,\mu}\xi_\alpha\xi_\beta\,,\qquad
J^\mu=g^{\mu\nu}\xi_\nu\,,
\ee
\be
{\cal C}_\mu=-\phi_{,\mu}
-{1\over 4}(\log g)_{,\mu}
+{\cal B}_\mu\,,\qquad
\bar{\cal C}_\mu
=-\phi_{,\mu}
-{1\over 4}(\log g)_{,\mu}
-{\cal B}_\mu\,,
\ee
\be
\alpha=-\xi_\mu g^{\mu\nu}{}_{,\nu}-2\xi_\mu g^{\mu\nu}{\cal B}_\nu\,,
\qquad
\bar\alpha=-\xi_\mu g^{\mu\nu}{}_{,\nu}+2\xi_\mu g^{\mu\nu}{\cal B}_\nu\,,
\ee
\be
\beta_\mu(\tau)=\tau H_\mu
=\tau g^{\alpha\beta}{}_{,\mu}\xi_\alpha\xi_\beta\,.
\ee

The
integrals over the cotangent vector $\xi$ are simply Gaussian
and can be easily computed. First of all, we 
have
\be
\int\limits_{\RR^n} {d\xi\over \pi^{n/2}}\,e^{-H}
=g^{1/2} \,.
\ee
So, we immediately obtain the coefficient $A_0$ as the Riemannian volume
of the manifold
\be
A_0=\int\limits_M dx\, g^{1/2}\,.
\ee
Next, we
introduce the notation
for the Gaussian averages
\be
\left<f\right>=\int\limits_{\RR^n} {d\xi\over \pi^{n/2}}\,g^{-1/2}e^{-H}
f(\xi)\,.
\ee
Then the Gaussian average of an exponential function gives
the generating function
\be
\left<\exp({\xi x})\right>
=\exp\left({{1\over 4}g_{\mu\nu}x^\mu x^\nu}\right)\,,
\ee
where $g_{\mu\nu}$ is the inverse matrix of the matrix
$g^{\mu\nu}$.
Expansion in the power series in $x$ generates the Gaussian
averages of polynomials
\bea
\left<\xi_{\mu_1}\cdots\xi_{\mu_{2n+1}}\right>
&=&0,
\\[10pt]
\left<\xi_{\mu_1}\cdots\xi_{\mu_{2n}}\right>
&=&
{(2n)!\over 2^{2n}n!}
g_{(\mu_1\mu_2}\cdots g_{\mu_{2n-1}\mu_{2n})},
\eea
where the parenthesis denote complete symmetrization over all
indices. In particular,
\be
\left<1\right>=1, 
\qquad 
\left<\xi_\mu\xi_\nu\right>={1\over 2}g_{\mu\nu},
\qquad
\ee
\bea
\left<\xi_\mu\xi_\nu\xi_\alpha\xi_\beta\right>
&=&{3\over 4}g_{(\mu\nu}g_{\alpha\beta)}
\nonumber\\
&=&{1\over 4}(g_{\mu\nu}g_{\alpha\beta}
+g_{\mu\alpha}g_{\nu\beta}
+g_{\mu\beta}g_{\nu\alpha})\,,
\eea
\bea
&&\left<\xi_\mu\xi_\nu\xi_\alpha\xi_\beta\xi_\rho\xi_\sigma\right>
={15\over 8}g_{(\mu\nu}g_{\alpha\beta}g_{\rho\sigma)}
\nonumber\\
&&={1\over 8}(
g_{\mu\nu}g_{\alpha\beta}g_{\rho\sigma}
+g_{\mu\alpha}g_{\nu\beta}g_{\rho\sigma}
+g_{\mu\beta}g_{\nu\alpha}g_{\rho\sigma}
+g_{\mu\rho}g_{\alpha\beta}g_{\nu\sigma}
+g_{\mu\sigma}g_{\alpha\beta}g_{\rho\nu}
\nonumber\\
&&
+g_{\mu\nu}g_{\alpha\rho}g_{\beta\sigma}
+g_{\mu\alpha}g_{\nu\rho}g_{\beta\sigma}
+g_{\mu\beta}g_{\nu\rho}g_{\alpha\sigma}
+g_{\mu\rho}g_{\alpha\nu}g_{\beta\sigma}
+g_{\mu\sigma}g_{\alpha\rho}g_{\beta\nu}
\nonumber\\
&&
+g_{\mu\nu}g_{\alpha\sigma}g_{\beta\rho}
+g_{\mu\alpha}g_{\nu\sigma}g_{\beta\rho}
+g_{\mu\beta}g_{\nu\sigma}g_{\alpha\rho}
+g_{\mu\rho}g_{\alpha\sigma}g_{\beta\nu}
+g_{\mu\sigma}g_{\alpha\nu}g_{\beta\rho}
)\,.
\nonumber\\
\eea

We have
\bea
A_1&=&\int\limits_M dx\, g^{1/2}\tr_V\,\Biggl<-q
-\int\limits_0^1 d\tau_1 
[\bar{\cal C}_\mu-(1-\tau_1)H_\mu]
g^{\mu\nu}
({\cal C}_\nu-\tau_1 H_\nu)
\nonumber\\
&&
+\int\limits_0^1d\tau_2\int\limits_0^{\tau_2}d\tau_1 
[\bar\alpha+2(1-\tau_2) h]
(\alpha+2 \tau_1 h)
\Biggr>
\,,
\eea
where
\be
h=J^\mu H_\mu
=g^{\mu\rho}g^{\alpha\beta}{}_{,\mu}\xi_\alpha\xi_\beta\xi_\rho\,.
\ee
Evaluating the integrals over the parameters $\tau_1$ and $\tau_2$
we obtain
\bea
A_1&=&\int\limits_M dx\, g^{1/2}\tr_V\,\Biggl<-q
+(g^{\mu\nu}-2J^\mu J^\nu){\cal B}_\mu{\cal B}_\nu
\nonumber\\
&&
-g^{\mu\nu}\phi_{,\mu}\phi_{,\nu}
-g^{\mu\nu}\left[H_\mu+{1\over 2}(\log g)_{,\mu}\right]\phi_{,\nu}
\nonumber\\
&&
-{1\over 4}g^{\mu\nu}H_\mu (\log g)_{,\nu}
-{1\over 16}g^{\mu\nu}(\log g)_{,\mu}(\log g)_{,\nu}
\nonumber\\[10pt]
&&
-{1\over 6}g^{\mu\nu}H_\mu H_\nu
+{1\over 2}\lambda^2
-{2\over 3}h\lambda
+{1\over 6}h^2
\Biggr>
\,,
\eea
where
\be
\lambda=\xi_\mu g^{\mu\nu}{}_{,\nu}\,.
\ee


Lastly, we need to evaluate the Gaussian averages. By taking into account
the formulas above we obtain
\bea
\left<J^\mu J^\nu\right>
&=&{1\over 2}g^{\mu\nu}\,,
\\
\left<H_\mu\right>
&=&{1\over 2}g_{\alpha\beta}g^{\alpha\beta}{}_{,\mu}
=-{1\over 2}(\log g)_{,\mu}\,,
\\
\left<H_\mu H_\nu\right>
&=&{1\over 4}(\log g)_{,\mu}(\log g)_{,\nu}
+{1\over 2}g_{\rho\alpha}g^{\alpha\beta}{}_{,\mu}
g_{\beta\sigma}g^{\sigma\rho}{}_{,\nu}\,,
\\
\left<\lambda^2\right>
&=&{1\over 2}g_{\mu\nu}g^{\mu\alpha}{}_{,\alpha}
g^{\nu\beta}{}_{,\beta}\,,
\\
\left<h\lambda\right>
&=&-{1\over 4}(\log g)_{,\mu}g^{\mu\alpha}{}_{,\alpha}
+{1\over 2}g_{\mu\nu}g^{\mu\alpha}{}_{,\alpha}
g^{\nu\beta}{}_{,\beta}\,,
\\
\left<h^2\right>
&=&{15\over 8}
g_{(\mu\nu}g_{\alpha\beta}g_{\rho\sigma)}
g^{\gamma\rho}g^{\delta\sigma}
g^{\mu\nu}{}_{,\gamma}
g^{\alpha\beta}{}_{,\delta}
\nonumber\\
&=&
{1\over 8}\Biggl[
g^{\gamma\delta}(\log g){,_\gamma}(\log g)_{,\delta}
-4g^{\mu\nu}{}_{,\mu}(\log g)_{,\nu}
\\
&&
+2g_{\mu\alpha}g_{\nu\beta}g^{\gamma\delta}
g^{\mu\nu}{}_{,\gamma}g^{\alpha\beta}{}_{,\delta}
+4g_{\mu\alpha}
g^{\mu\nu}{}_{,\nu}g^{\alpha\beta}{}_{,\beta}
+4g_{\beta\nu}
g^{\mu\nu}{}_{,\alpha}g^{\alpha\beta}{}_{,\mu}
\Biggr]
\,.
\nonumber
\eea
Here we used the equation
\be
(\log g)_{,\mu}=-g_{\alpha\beta}g^{\alpha\beta}{}_{,\mu}
=g^{\alpha\beta}g_{\alpha\beta,\mu}
\ee

We see that $A_1$ does not depend
on ${\cal B}$ at all and we get finally
\bea
A_1&=&\int\limits_M dx\, g^{1/2}N\,\Biggl\{-q
-g^{\mu\nu}\phi_{,\mu}\phi_{,\nu}
\nonumber\\
&&
+{1\over 24}g^{\mu\nu}(\log g)_{,\mu}(\log g)_{,\nu}
+{1\over 12}(\log g)_{,\mu}g^{\mu\nu}{}_{,\nu}
\nonumber\\[10pt]
&&
+{1\over 12}g_{\beta\nu}g^{\mu\nu}{}_{,\alpha}g^{\alpha\beta}{}_{,\mu}
-{1\over 24}g_{\mu\alpha}g_{\nu\beta}g^{\gamma\delta}
g^{\mu\nu}{}_{,\gamma}g^{\alpha\beta}{}_{,\delta}
\Biggr\}
\,.
\eea

One can further integrate by parts here but we will stop here.
Since the part that depends only on $g$
must be an invariant, it must be an integral of a scalar
quantity, which can be only the scalar curvature (there is
only one such scalar). It turns out that
\bea
A_1&=&\int\limits_M dx\, g^{1/2}N\,\left(-q
-g^{\mu\nu}\phi_{,\mu}\phi_{,\nu}
+{1\over 6}R\right)
\nonumber\\
&=&
\int\limits_M dx\, g^{1/2}N\,\left(-\tilde q
+{1\over 6}R\right)
\,,
\eea
where $R$ is the scalar curvature of the metric $g$
and $\tilde q$ is defined by (\ref{194}).
This coincides with the standard results 
\cite{gilkey95,avramidi99,avramidi02}
for the Laplace operator.

\section{Noncommutative Gauged Gravity}

Our main idea is that gravity has some {\it new degrees of
freedom}. Instead of describing gravity by one tensor field we propose
to describe it by new dynamical variables: the matrix valued tensor
field $a^{\mu\nu}$ and the density matrix field $\rho$. 

\subsection{Matter in Gravitational Field}

%
%

The motion of a massive
particle in the gravitational field should be determined 
by the principle of the extremal action. In general relativity the 
action is simply the interval $S=-m\int \sqrt{-ds^2}$, 
which means that particles
propagate along the geodesics of the metric $g_{\mu\nu}$. The parameter
$m$ is the mass of the particle. In our modified theory, we assume that
a particle ``splits'' into several parts which propagate separately and
have different masses. In other words, we propose to describe the motion
of a particle in gravitational field by the action
\be
S_{\rm particle}= -\sum_{i=1}^s m_{(i)}\int \sqrt{-ds^2_{(i)}}\,,
\ee
where the intervals $ds^2_{(i)}$ are defined by (\ref{33}). 
In the commutative limit the sum of the mass parameters determines
the usual mass
\be
m=\sum_{i=1}^s m_{(i)}\,.
\ee
One can develop the whole Hamilton-Jacobi formalism starting from
this for the particle mechanics based on Finsler geometry. We will
not do it here but refer the reader to Rund's book \cite{rund59}
(in particular, Chap. 7 of the Russian edition).

The dynamics of matter fields in gravitational field can be now 
described by the differential calculus developed in Sect. 3.
A typical action for the matter fields has the form
\be
S_{\rm field}=\int\limits_M dx\left\{ -\left<\varphi,L\varphi\right>
+W(\varphi)\right\}\,,
\ee
where $L$ is an appropriate second-order differential operator
constructed by the methods of Sect. 3 and $W$ is a potential. 
The first-order operators 
of Dirac type needed for spinor fields can be constructed 
similarly. 

\subsection{Equations of Gravitational Field}

Our goal is to construct an action functional that is invariant under
both the diffeomorphisms and the local gauge transformations. We also
require that it is a noncommutative deformation of the Einstein general
relativity, which means that our action should reduce to the standard
Einstein-Hilbert functional in the commutative limit.

We construct the classical action following the ideology of ``induced
gravity'' as the large mass limit of the effective action. The classical
action of gravity is then identified with the first two coefficients of
the asymptotic expansion of the effective action as $m\to \infty$
(\ref{153}). In
other words, we use the coefficients $A_0$ and $A_1$  to construct the
invariant action functional of the noncommutative gravity. For a
dynamical theory we need an invariant action functional that depends on
the first derivatives of dynamical degrees of freedom. Such an invariant
is given by the heat kernel coefficient $A_1$. Therefore, the invariant
action functional of noncommutative gravity is a linear combination of
the first two heat kernel coefficients
\be
S={1\over 16\pi G}{1\over N}\left[6\, A_1-2\Lambda\,A_0\right]
+S_{\rm matter}\,,
\label{64}
\ee
where $S_{\rm matter}$ is the action of matter fields $G$ and
$\Lambda$ are the phenomenological  coupling constants (Newton constant
and the cosmological constant), and  the coefficients are chosen in such
a way that it has the correct commutative limit.
This action is in some
sort unique and the dynamical model described by it can be called the
{\it Induced Noncommutative  Gauged Gravity}. 

It is worth indicating the general form of the action. It can be represented
symbolically as follows
\bea
S&=&{1\over 16\pi GN}\int\limits_M dx \tr_V\Big\{
F(a,\rho)\partial a \partial a
+ F(a,\rho)\partial a \partial \rho
+F(a,\rho) \partial \rho\partial \rho
\nonumber\\
&&
\qquad\qquad
+F(a)(6Q+2\Lambda)
+F(a,\rho) \partial {\cal B}
+F(a,\rho) {\cal B}{\cal B}
\Big\}+S_{\rm matter}\,,
\label{65a}
\eea
where $\partial$ denotes the derivatives and $F(a,\rho)$
denotes the coefficients that can only depend on $a$ and $\rho$.
Of course, all the coefficients are different and one has to include
similar terms with all possible orderings of noncommutative factors.

The fundamental equations
of this model are
\be
{\delta S\over \delta a^{\mu\nu}}=0\,, \qquad
{\delta S\over \delta \rho}=0\,,\qquad
{\delta S\over \delta {\cal B}_\mu}=0\,.
\ee
Notice that the gravitational action does 
not depend on the derivatives of the vector field 
${\cal B}_\mu$ (after integration by parts). Therefore, if 
the matter action $S_{\rm matter}$ does not include
a kinetic term for the fields ${\cal B}$, then 
the variation with respect to ${\cal B}$ gives just
a constraint, which simply expresses ${\cal B}$ in 
terms of the derivatives of
the matrices $a^{\mu\nu}$ and $\rho$.
Also, the matrix $Q$ is fixed as a given self-adjoint positive definite
function of $a$ and $\rho$ and their derivatives.
Its form can be adjusted by imposing some additional
physical requirements.
This could be important for spontaneous symmetry breakdown.
For simplicity, it can be just set to zero $Q=0$.

%
%
%
%
%
%
%
%
%
%


%
%
%
%
%
%
%
%
%
%
%
%
%

If we introduce a deformation parameter $\varkappa$ according to
(\ref{547}):
$a^{\mu\nu}=g^{\mu\nu}\II+\varkappa h^{\mu\nu}$,
$\rho=g^{1/4}\exp(\II\phi+\varkappa \psi)$,
and $Q=q\II+\varkappa P$,
where $h^{\mu\nu}$ is a matrix-valued tensor field,
$\phi$ and $q$ are scalar fields, and $\psi$ and $P$ are
matrix-valued scalar fields,
then in the limit $\varkappa\to 0$ the action becomes
\be
S_{(0)}={1\over 16\pi G}\int\limits_M dx\,g^{1/2}\left(R-2\Lambda
-6g^{\mu\nu}\phi_{,\mu}\phi_{,\nu}-6q\right)
+S_{{\rm matter} (0)}\,,
\label{65}
\ee
which describes the Einstein general relativity and a scalar field.
Note that in the case $q=-g^{\mu\nu}\phi_{,\mu}\phi_{,\nu}-\Box\phi$
the scalar field disappears. Other interesting choices of $q$ are
$q=m^2$ or $q=\Lambda(\phi^2-m^2)^2$.

It would be certainly very interesting to compute the
deformation corrections to the action (\ref{65}), that is to 
understand the precise way under which the commutative action
(\ref{65}) gets deformed to the non-commutative one (\ref{64}).
In general, we should obtain
\be
S(a,\rho,{\cal B})=S_{(0)}(g,\phi)
+\sum_{k=1}^\infty \varkappa^k S_{(k)}(g,\phi,h,\psi,{\cal B})\,.
\ee
Notice that the zero-order action $S_{(0)}$ does not depend on the
fields ${\cal B}$. By a simple rescaling $\phi\to \varkappa \phi$ and
$q\to\varkappa q$ we could easily shift the dependence on these fields
to the higher orders as well so that the zero order term $S_{(0)}(g)$
is just the
Einstein-Hilbert action. The coefficients $S_{(k)}$ are polynomial in
the fields $\phi, h, {\cal B}$ and $\psi$. Their general form can be
read off from (\ref{65a}). In principle, one could get the coefficients
$S_{(k)}$ from the general result of Theorem 1. However, such a
calculation (even if straightforward) presents a real technical
challenge and would require a separate paper.  We plan to carry out this
calculation in the near future and present the results elsewhere.

\section{Discussion}

In conclusion we list some interesting open problems in the proposed model.

{\it Uniqueness of the noncommutative deformation.} First of all, it is
very important to understand whether the proposed deformation of the
general relativity is unique. If it is not unique, then what additional
physical conditions should one impose to make such a deformation unique.

{\it Interaction of noncommutative gravity with matter.} One needs to
find a consistent  way to describe the interaction of the ordinary
matter with gravity. In particular, to understand which physical matter
fields should interact with the noncommutative (gravicolor) degrees of
freedom and which should only feel the graviwhite part.

{\it Classical solutions and singularities.} It would be very
interesting problem to study simple solutions of this model, say a
static spherically symmetric solution, which would describe  a
``noncommutative black hole" as well as a time-dependent homogeneous
solution, which would describe the ``noncommutative cosmology''. One
should study the  problem of singularities of classical solutions. If
this model is free from singularities, it would be a very significant 
argument in
favor of it.

{\it  Spontaneous breakdown of symmetry.} One needs to understand
whether it is possible to introduce the spontaneous breakdown of the
gauge symmetry, so that in the broken phase in the vacuum there is just
one tensor field, which is identified with the metric of the space-time.
All other tensor fields must have zero vacuum expectation values. In the
unbroken phase there will not be a metric at all in the usual sense
since there is no preferred tensor field with non-zero vacuum
expectation value.

{\it Quantization and renormalization.}  Our model is, in fact, nothing
but a generalized sigma model. So, the problems in quantization of this
model are the same as in the quantization of the sigma model.

{\it Semi-classical (one-loop) approximation  and heat kernel
asymptotics.} We point out that the study of the one-loop approximation
requires new calculational methods since the partial differential
operators involved are not of the so-called Laplace type (nonscalar
leading symbol).  For example, even the heat
kernel coefficients $A_2$ needed for the
renormalization in four dimensions is not known in general.

{\it High-energy behavior.} We expect that the behavior of our model at higher
energies should be radically different from the Einstein gravity since
there is no preferred metric in the unbroken phase, when the new gauge
symmetry is intact.

{\it Dark energy.} It would be very interesting to study the question
whether the new noncommutative degrees of freedom of gravity could be
accounted for the dark energy in cosmology.

{\it Low energy behavior and confinement.} One could
expect the gauge ({\it gravicolor}) degrees of freedom to be confined
within some short characteristic scales (say, Planck scale), 
so that only the invariants ({\it graviwhite}
states) are visible at large distances. Then the metric and the curvature
would be only effective characterictics of the spacetime at large distances.

{\it Noncommutative deformation of Riemannian geometry.} The Einstein
spaces are manifolds with the metric satisfying the vacuum Einstein
equations (with cosmological constant), i.e. $R_{\mu\nu}=\Lambda
g_{\mu\nu}$ with some constant $\Lambda$. In other words, the Einstein
metrics are the extremals of the Einstein-Hilbert functional. The study
of Einstein spaces is a very important subject in Riemannian geometry.
It would be very interesting to study the noncommutative
deformations of Einstein spaces defined as the extremals of the
functional of matrix gravity.
This would also have deep connections to
``{\it noncommutative spectral geometry}''.


\end{document}